\documentclass[
amsmath,amssymb,
aps,pra,
showpacs,
preprint, 
superscriptaddress,
longbibliography
]{revtex4-1}

\usepackage{pdfpages}
\usepackage{graphicx}
\usepackage{bm}
\usepackage{color}
\usepackage[linkcolor=red, citecolor=black, urlcolor=blue, colorlinks=true]{hyperref}
\makeatletter
\newcommand{\colorcaption}[2][]{
  \begingroup
  \renewcommand{\@caption@fignum@sep}{ (color online). } 
  \caption[#1]{#2}
  \endgroup}
\makeatother

\begin{document}

\title{Stationary light pulses and narrowband light storage in a laser-cooled ensemble loaded into a hollow-core fiber}

\author{Frank Blatt}
\affiliation{Institut f\"ur Angewandte Physik, Technische Universit\"at Darmstadt, Hochschulstra{\ss}e 6, 64289 Darmstadt, Germany}

\author{Lachezar S. Simeonov}
\affiliation{Department of Physics, St. Kliment Ohridski University of Sofia, 5 James Bourchier Boulevard, 1164 Sofia, Bulgaria}

\author{Thomas Halfmann}
\affiliation{Institut f\"ur Angewandte Physik, Technische Universit\"at Darmstadt, Hochschulstra{\ss}e 6, 64289 Darmstadt, Germany}

\author{Thorsten Peters}
\affiliation{Institut f\"ur Angewandte Physik, Technische Universit\"at Darmstadt, Hochschulstra{\ss}e 6, 64289 Darmstadt, Germany}

\date{\today}

\begin{abstract}
We report on the first observation of stationary light pulses and narrowband light storage inside a hollow-core photonic crystal fiber. Laser-cooled atoms were first loaded into the fiber core providing strong light-matter coupling. Light pulses were then stored in a collective atomic excitation using a single control laser beam. By applying a second counterpropagating control beam, a light pulse could be brought to a standstill. Our work paves the way towards the creation of strongly-correlated many-body systems with photons and applications in the field of quantum information processing.
\end{abstract}

\pacs{
37.10.Gh,   		
32.80.Qk, 		  	
03.67.-a,					
42.50.Gy,   	 	
}																			
\maketitle

\textit{Introduction.}---Achieving strong coupling of light and matter is a long pursued goal in the field of quantum optics. Not only does it allow for large linear light storage efficiencies \cite{GAL07b}, 
e.g., using electromagnetically induced transparency (EIT) \cite{FIM05,CLW13,SHL16} or a gradient-echo technique \cite{HLA08,CCE16}. 
It also sets the ground for quantum nonlinear optics (NLO), where strong interactions between \textit{individual} photons can be mediated via coupling through the medium \cite{CVL14}. 
This would, e.g., enable all-optical quantum networks \cite{Kimble2008}, 
the creation of strongly-correlated light-matter systems \cite{CGM08,KH10b,AHK11}, 
single-photon switches \cite{HCG12}, 
or the simulation of relativistic theories with photons \cite{AHC13}. 

\begin{figure}[bp]
\includegraphics[width=12cm]{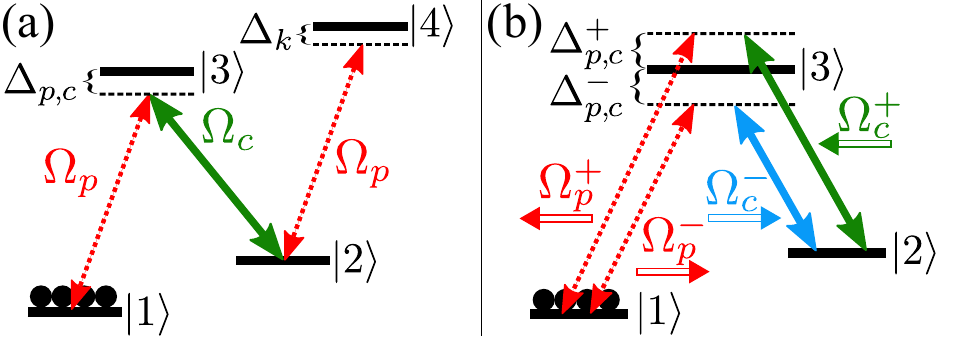}
\colorcaption{\label{fig:Fig1} Coupling schemes for EIT with a Kerr-type nonlinearity (a) and SLPs (b). $\Omega_{p,c}$ are the Rabi frequencies of the probe/control fields.}
\end{figure}

Strong NLO interactions at the single-photon level have already been demonstrated, e.g., using long-range atom-atom interactions \cite{PFL12,DK12,FPL13,PWA13} 
or placing atoms into high-finesse cavities \cite{Birnbaum2005a,Dayan2008}.
However, NLO interactions as proposed in \cite{CGM08,KH10b,AHK11,HCG12,AHC13} 
using atomic ensembles coupled to one-dimensional (1D) waveguides are still waiting for their experimental realization, as they rely on the effect of stationary light pulses (SLPs), i.e., light pulses with a quasi-stationary envelope \cite{BZL03}, created in an atomic ensemble with strong light-matter coupling.
The coupling strength of light to an ensemble of $N_{\mathrm{atom}}$ atoms for \textit{linear} interactions, such as light storage in a collective atomic excitation, is determined by the optical depth $\mathrm{OD}$$\,=\,$$-\ln{T}$, with the resonant transmission $T$ \cite{GAL07b}.
For \textit{nonlinear} interactions the simultaneous interaction of an atom with multiple photons is required. Here, the relevant quantity is the product of the OD and the OD per atom \cite{HCG12}. The OD per atom is given by $\mathrm{OD}^*$=$\mathrm{OD}/N_{\mathrm{atom}}$$\approx$$\sigma_a/A_w$, i.e., the probability of an atom (absorption cross section $\sigma_a$) to interact with a photon of the guided mode (mode  area $A_w$) \cite{CVL14}. Impressive results have been achieved with atoms coupled to photonic nanowaveguides \cite{GHY14,GHH15} and nanofibers \cite{VRS10,GCA12}. 
On the other hand, hollow-core photonic crystal fibers (HCPCFs) loaded with atoms \cite{GBR06,CWS08,VMW10,BHP11}
 provide a smaller $\mathrm{OD}^*$ due to a larger mode area but allow for significantly larger ODs \cite{GBR06,BHP11,BHP14,KSS15}.
Once an ensemble with strong light-matter coupling is provided, SLPs with a Kerr-type nonlinearity \cite{SI96,HH99,HCG12} (see Fig.~\ref{fig:Fig1}) have to be created in order to achieve strong NLO interactions. 
The effect of EIT \cite{FIM05} can be illustrated by the coupling scheme shown in Fig.~\ref{fig:Fig1}(a) without the coupling to state $|4\rangle$. 
In this $\Lambda$-type system the strong control field renders an opaque medium transparent for the weak (copropagating) probe field due to quantum interference near two-photon resonance at $\Delta_p$=$\Delta_c$. Dark-state polaritons (DSPs) \cite{FL00} 
are formed whose group velocity can be controlled by the Rabi frequency $\Omega_c$ of the control \cite{FL02}. 
By adiabatically switching off the control while the DSPs are created, the probe field is mapped onto a non-propagating long-lived collective atomic excitation. It can be retrieved by reapplying the control. This is termed light storage and retrieval \cite{LDB01} 
and has led, e.g., to impressive light storage times \cite{HHH13}. 
As no light is present inside the medium during the storage period, NLO interactions are not possible.
However, when a second counterpropagating control beam is added (see Fig.~\ref{fig:Fig1}(b)) while the DSPs propagate through the medium, an all-optical cavity is created. The DSPs are effectively stopped with a non-vanishing photonic component and a quasi-stationary envelope \cite{BZL03,LLP09}, i.e., light pulses are trapped without a physical cavity. 
Although the envelope is quasi-stationary, the DSPs within the pulse still jitter back and forth with finite group velocity. With a Kerr-type coupling to a state $|4\rangle$ being present this results in (in)elastic collisions of the DSPs, depending on the magnitude of the detunings $\Delta_{p,k}$ \cite{HCG12}. 
Thus, NLO interactions using SLPs become possible \cite{ABZ05,CLH12}. 

Towards the goal of implementing efficient fiber-based linear and NLO interactions at the quantum level, in our work we present now the experimental implementation of narrowband EIT in a cold 1D ensemble with large optical depth up to $\mathrm{OD}$=$400$ (i.e. two orders of magnitude larger than in previous experiments \cite{BHB09}),
the first demonstration of light storage and retrieval, and the first observation of SLPs in a HCPCF loaded with cold atoms. We compare our results to elaborated numerical simulations. Finally we discuss strategies towards future applications.

\begin{figure}
\includegraphics[width=12cm]{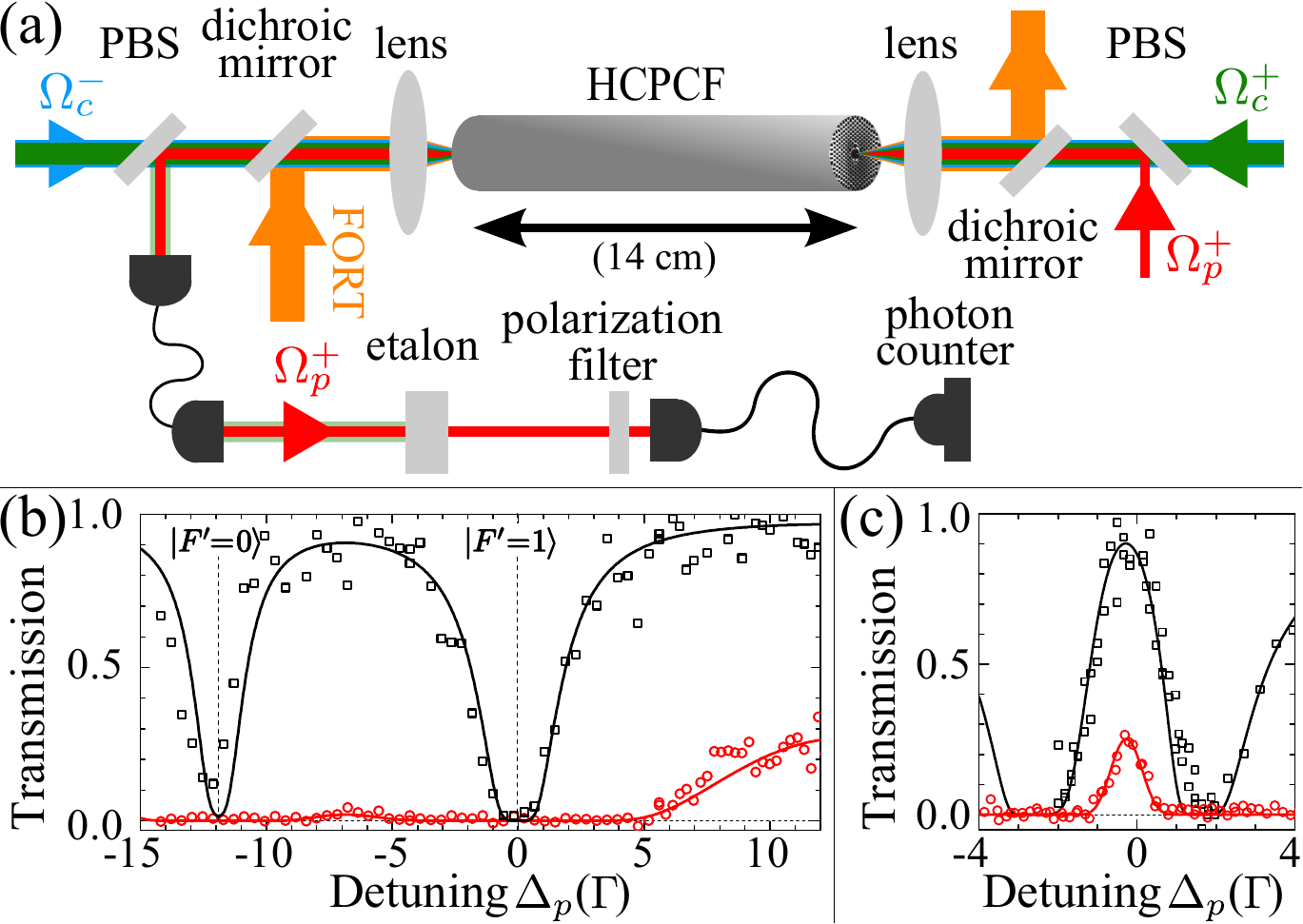}
\colorcaption{\label{fig:setup_EIT}(a) Schematic experimental setup. PBS: polarizing beam splitter. (b) Experimental (symbols) and simulated (lines) transmission through the HCPCF filled with atoms for ODs of 20 (black squares) and 400 (red circles) without control field. (c) EIT at resonance for the parameters OD=20, $\Omega_c^+$=4.5$\Gamma$, $\Delta_c^+$=0.7$\Gamma$, $\Theta$=550(50)$\,\mu$K (black squares) and OD=400, $\Omega_c^+$=6.1$\Gamma$, $\Delta_c^+$=1.8$\Gamma$, $\Theta$=450(50)$\,\mu$K (red circles).}
\end{figure}

\textit{Experiment.}---A schematic overview of the experimental setup is shown in Fig.~\ref{fig:setup_EIT}(a). We loaded laser-cooled $^{87}$Rb atoms from a magneto-optical trap (MOT) into the core of a vertically aligned HCPCF (HC-800-02, NKT Photonics, core diameter $\sim$7~$\mu$m) \cite{BHP14} 
at a rate of 0.76~Hz.
A nearly Gaussian-shaped red-detuned far-off-resonant trap (FORT) \cite{GWO00} 
(trap depth $\sim$5~mK inside the HCPCF) prevented collisions of the cold atoms with the room-temperature fiber wall, allowing for guiding and tight confinement of the atoms.
The total number of atoms $N_{\mathrm{atom}}$$\lesssim$$10^5$
loaded into the fiber was controlled by the power of the repumper beam of the MOT.
For a 5~mK deep FORT a temperature around 450~$\mu$K of the atoms inside the HCPCF can be expected. However, due to an unresolved heating mechanism, the temperature varied day by day between $350~\mu$K$\leq$$\Theta$$\leq$1.1~mK \cite{BHP11,BHP14}. 
To avoid inhomogeneous broadening by the deep FORT, we modulated the trap depth with an on-off ratio of $\geq$$14~$dB and $\omega_{\mathrm{mod}}$$=$$2\pi$$\times$$250~$kHz. 
This provided up to 50 measurement periods $\tau_{\mathrm{meas}}$$\leq$$3~\mu$s 
each, with insignificant losses and one-photon transition shifts $<$$0.2\Gamma$ ($\Gamma$=2$\pi\times$6.07~MHz is the excited state linewidth).

We implemented the $\Lambda$ scheme for EIT as shown in Fig.~\ref{fig:Fig1}(b) with $|1\rangle$=$|5^2\mathrm{S}_{1/2},F$=1$\rangle$, $|2\rangle$=$|5^2\mathrm{S}_{1/2},F$=2$\rangle$, and $|3\rangle$=$|5^2\mathrm{P}_{3/2},F'$=1$\rangle$. With these couplings EIT is achieved for all possible transitions between Zeeman levels for an arbitrary polarization of the laser fields due to the birefringence typical for HCPCFs \cite{SMU05,CLV04}. Two external-cavity diode lasers, locked \cite{BHW07} 
with a relative bandwidth of 8~kHz during an integration period of 3.5~s provided the probe and control fields. 
Magnetic stray fields were compensated by a 3D magnetic offset field which had to be adjusted slightly day by day.

The weak probe field (3.5~pW $\leq P_p \leq$ 700~pW) was filtered from the much stronger and exactly collinear forward control field (50~nW $\leq P_c\leq$ 1.9~$\mu$W) 
after the HCPCF for detection with a photon counter (PerkinElmer, SPCM AQRH-12) by a monolithic etalon \cite{PML12} 
combined with polarization and spatial filtering \cite{BHP11}. 
The detection efficiency of the probe was 10~\% 
and the relative attenuation of the control was 64~dB. 
We always kept the probe power low enough to fulfill $\Omega_p$$\ll$$\Omega_c$ and to have much smaller densities of the probe photons than the atoms, as required for DSP propagation \cite{FL02}. 
We counted the output pulses of the SPCM either directly with a counterboard (NI, PCI-6602) or with a digital storage oscilloscope (Agilent, DSO1014A) followed by software analysis. 
For further details on the experimental setup we refer the reader to \cite{BHP14} 
and the Supplemental Material \cite{SM}.

\textit{Simulations.}---
In order to compare our measurements to theoretical predictions and to characterize our setup we set up several numerical simulations. Due to the changing elliptical light polarization inside the HCPCF and a potentially small remaining magnetic field, we chose to model our experiments assuming an isotropic polarization using effective dipole moments [see Eqn.~(43) in \cite{Steck2015}] in all simulations.
We extended the theoretical models typically used to simulate the transmission through a fiber loaded with atoms \cite{BHP11} 
and pulse propagation in a medium driven by EIT with counterpropagating control fields \cite{ZAL06,WAL10b,PSC12} to incorporate the off-resonant states $|5^2\mathrm{P}_{3/2},F'$=0,2,3$\rangle$.
This is necessary as for $\mathrm{OD}$$\gtrsim$130 
the resonances of the D$_2$ line start to overlap. Also, inhomogeneous broadening of the two-photon resonance due to the radially inhomogeneous control fields $\Omega_c(r)$ and the atomic density profile $n(r)$ determined by the FORT potential and the temperature $\Theta$ \cite{GWO00} have to be considered. 
The transmission is then given by
\begin{equation}
T(\Delta_p,\Theta)\!=\!\exp\!\left[-\Gamma\frac{\sigma_a}{\sigma_{pc}^2} L\!\int\!\!\!\!\int\!\!n(r) p(v) \Omega_p(r)\alpha(\Delta_p) r \; dr dv\right] \notag
\end{equation}
with the measured mode field $^1\!\!/\!e^2$ radius $\sigma_{pc}$ of the probe/control beams, the normalized probe field profile $\Omega_p(r)$ and the Maxwell-Boltzmann distribution $p(v)$ for the temperature $\Theta$. The absorption coefficient $\alpha(\Delta_p)$ was taken from \cite{PWB12} 
and adjusted to the current coupling scheme while state $|5^2\mathrm{P}_{3/2},F'$=3$\rangle$ was adiabatically eliminated (see Supplemental Material~\cite{SM} for further details). We assumed the ground state decoherence rate $\gamma_{21}$=$\gamma_{\mathrm{trd}}$ to be dominated by transit relaxation decay \cite{SNH96}. 
Only the number of atoms $N_{\mathrm{atom}}$ inside the HCPCF and $\Theta$ were free parameters. The other parameters were initially chosen according to the measurements and were then allowed to be changed within the experimental uncertainties to best reproduce the measured data.

A comparison of the transmission spectra for homogeneous and inhomogeneous laser fields and medium, respectively, shows that the inhomogeneities can be included in the (homogeneous) 1D simulation by using effective control Rabi frequencies $\Omega_c^{\mathrm{eff}}=\beta\Omega_c$ and an additional decoherence rate $\gamma_{\mathrm{inh}}$$\propto$$\Omega_c^2,\Theta^2$ due to inhomogeneous broadening.
For the values of $\Theta$ and $\Omega_c$ used in our experiment we have $\gamma_{\mathrm{inh}}$$>$$\gamma_{\mathrm{trd}}$.
Although this allows one in general to discuss the experimental data in terms of EIT window width $\Delta\omega_{\mathrm{EIT}}=(\Omega_c^{\mathrm{eff}})^2/\Gamma\sqrt{\mathrm{OD}}$ compared to the decoherence rate $\gamma_{21}$$\approx$$\gamma_{\mathrm{inh}}$, things are more complicated due to incoherent absorption by state $|5^2\mathrm{P}_{3/2},F'$=0$\rangle$ in our multi-level system.
For simulating the pulse propagation through the medium under EIT conditions we either used a convolution of the probe pulse and the spectral transmission function $T(\Delta_p,\Theta)$ in 3D, or we solved the 1D Maxwell-Bloch equations \cite{WAL10b,PSC12} 
using $\Omega_c^{\mathrm{eff}}$ and $\gamma_{\mathrm{inh}}$ \cite{SM}.
The latter method was also used to simulate light storage and SLPs.

\textit{Results \& Discussion.}---
Figure~\ref{fig:setup_EIT}(b) shows the transmission through the HCPCF as a function of probe laser detuning for two different ODs. Each data point corresponds to $N_{\mathrm{avg}}$=$50$~averages 
measured during a gate time of $\tau_g$$=$$680$~ns. 
In the absence of EIT the medium can be rendered highly opaque over a broad frequency range depending on the number of atoms loaded into the fiber. By switching on the control beam during the measurement for the same conditions, the typical 
transmission window of EIT appears [Fig.~\ref{fig:setup_EIT}(c)]. The detuning $\Delta_c^+$$>$$0$ was here adjusted to compensate the two-photon ac Stark shift. The temperature of the medium was $\Theta$$=$$500(100)~\mu$K [$\gamma_{\mathrm{trd}}$$=$$0.008(1)\Gamma$] 
according to the simulations. This is in agreement with the expected lower temperature limit for the $\sim$5~mK deep FORT. 
Whereas we observed almost complete transmission for a moderate OD (black squares), 
the maximum transmission reaches only 25~\% 
for a high OD (red circles). 
According to our simulation, 30~\% of the absorption are due to incoherent absorption by state $|5^2\mathrm{P}_{3/2},F'$=$0\rangle$ which is not coupled by the control due to selection rules. 
We therefore chose $\mathrm{OD}$$\lesssim$140 for all following experiments to avoid a resonance overlap.

\begin{figure}
\includegraphics[width=12cm]{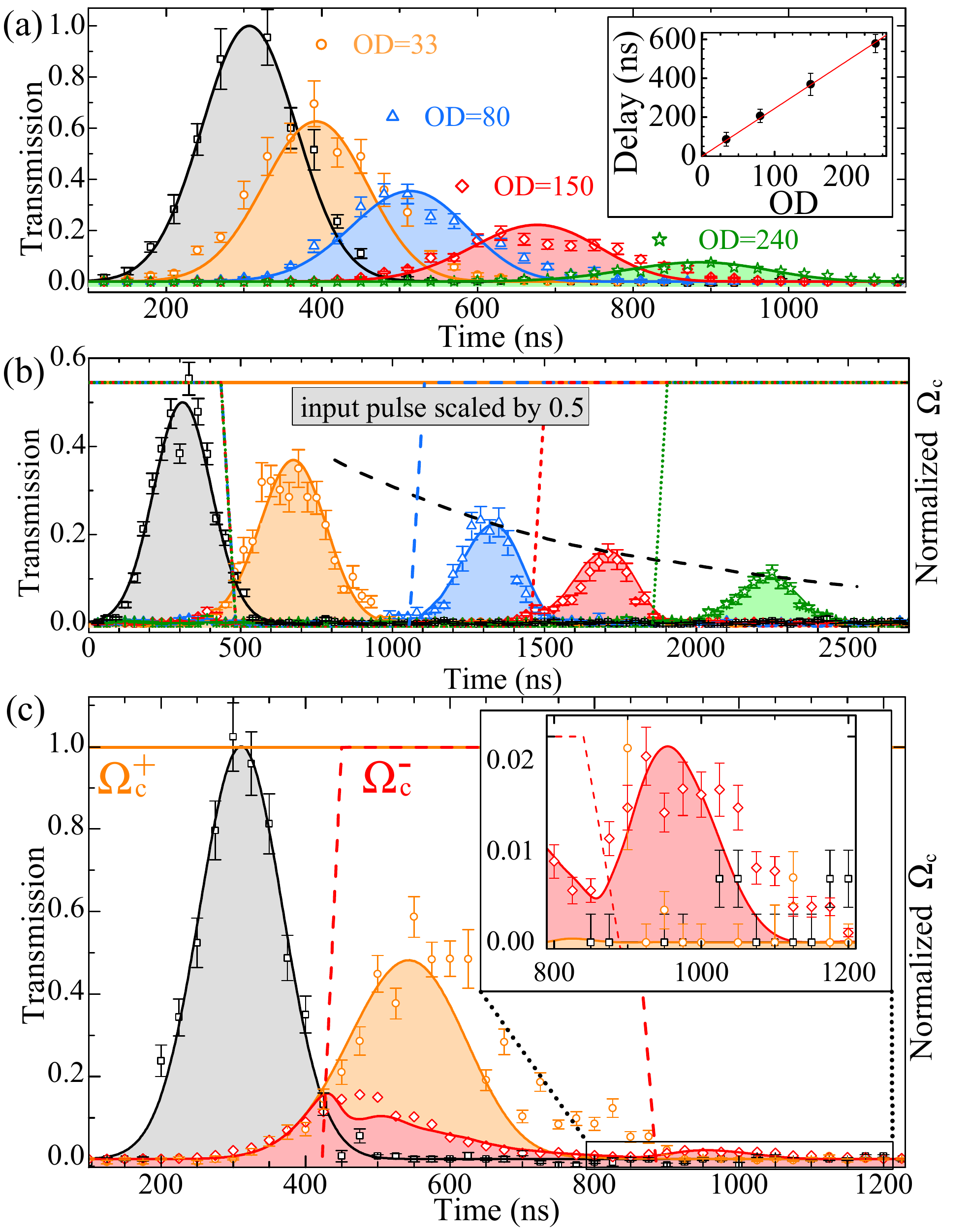}
\colorcaption{\label{fig:SL_LS_SLP}Normalized transmission of a Gaussian input probe pulse (black squares) through the HCPCF. Symbols depict experimental data and lines simulations. The control Rabi frequencies are indicated by line segments. The timescales are different in all plots.
(a) Slow light: The input pulse is delayed depending on the OD for $\Omega_c^+$=$3.8\Gamma$. $\Theta$=575(75)$\,\mu$K, $\gamma_{21}$=0.037(3)$\Gamma$. 
(b) Light storage: The input pulse is delayed by approximately one pulse width for constant $\Omega_c^+$ (orange circles and line). The solid, dashed, and dotted lines represent $\Omega_c^+(t)$ with colors according to the respective transmission for different storage times of 0.6~$\mu$s (blue triangles) and 1~$\mu$s (red diamonds) with OD=145(5), and a time of 1.4~$\mu$s (green stars) with OD=195(5). $\Omega_c$=3.7$\Gamma$, $\Theta$=450(50)$\,\mu$K, $\gamma_{\mathrm{trd}}$=0.008(1)$\Gamma$. 
(c) Slow light: $\Omega_c^+$=2.6$\Gamma$, $\Omega_c^-$=0, $\gamma_{\mathrm{inh}}$=$0.003(1)\Gamma$ (orange circles). SLP: $\Omega_c^+$=$2.6\Gamma$, $\Omega_c^-$=$3.8\Gamma$, $\gamma_{\mathrm{inh}}$=$0.012(1)\Gamma$ (red diamonds). OD=53, $\Theta$=350(50)$\,\mu$K, $\gamma_{\mathrm{trd}}$=$0.006(1)\Gamma$, $\Delta_c^+$=$+1.0\Gamma$, $\Delta_p^+$=$+0.45\Gamma$, $\Delta_c^-$=$-2.5\Gamma$. The inset shows an enlarged version for $t$$>$800~ns.}
\end{figure}

Using the steep dispersion within the EIT transparency window, the group velocity $v_g$ of a probe pulse can be significantly reduced as compared to the vacuum value $c$ \cite{FIM05}. 
As the pulse is simultaneously spatially compressed by $v_g/c$, it can be stored efficiently inside a medium much shorter than the original pulse length. The temporal delay can be estimated by $\tau_d=\Gamma \, \mathrm{OD}/\Omega_c^2$ \cite{FL02}. 
With the experimentally determined $\Omega_c$ (measuring the transmitted power and considering the losses of the optics) this allows one to determine the OD independently from the transmission spectra. Both results agree well within the experimental uncertainties.
In Fig.~\ref{fig:SL_LS_SLP}(a) we show the delays of an incident probe pulse (black squares) for constant $\Omega_c^+$ and varying OD ($N_{\mathrm{avg}}$=$1250$, $\tau_g$=$60$~ns).
The delay increases linearly with the OD in accordance with the theoretical expectation (see inset).
The probe pulse can be delayed by more than one pulse length, i.e., it can be compressed such that it completely fits inside the medium.
We note that only by including the radially inhomogeneous profiles $\Omega_{p,c}(r)$ and $n(r)$, the simulations show good agreement with the measurements for realistic parameters. 

By adiabatically switching off the control field while the probe pulse is inside the medium, we can now map the probe pulse onto a collective atomic excitation \cite{LDB01,FIM05}. 
Switching on the same control field after a certain storage time leads to a retrieval of the probe field into its original direction. 
Figure~\ref{fig:SL_LS_SLP}(b) shows the incident probe pulse (black squares), the delayed pulse without storage (orange circles), and the retrieved pulses for different storage times $\Delta\tau$ (blue triangles, red diamonds and green stars; $N_{\mathrm{avg}}$=$1250$, $\tau_g$=$60$~ns).
The probe pulse is attenuated by 63~\% 
as it moves through the medium (orange circles). This attenuation cannot be solely attributed to decoherence by transit relaxation decay, but requires the inclusion of inhomogeneous broadening [$\gamma_{\mathrm{inh}}$=$0.015(1)\Gamma$] by the control while $\Omega_c^+(t)$$\neq$0 in the effective 1D simulation. 
Once the pulse is temporarily stored [$\Omega_c^+(t)$=0], it decays exponentially (black dashed line) with a decoherence rate $\gamma_{21}$=$0.009(1)\Gamma$ dominated by transit relaxation decay ($\gamma_{\mathrm{trd}}$=$0.008\Gamma$) for a temperature of $450(50)~\mu$K. We note that this exponential decay was achieved after careful suppression of the stray magnetic field \cite{PCW09}. Before, collapses and revivals as in \cite{GMD15} were observed. 
The efficiency, defined as output/input pulse area, is $\eta$=$23(5)~\%$ for 0.6~$\mu$s of storage.   
This is about 2$\times$ (8$\times$) larger than for similar measurements performed with nanofibers \cite{SCA15,GMD15} 
(with similar decoherence rates) due to the larger OD but also larger inhomogeneous broadening. The latter one is also responsible for the lower efficiency as expected for free-space setups of comparable OD \cite{CLW13}.
Applying a cooling technique \cite{WDB99} inside the fiber (to reduce inhomogeneous broadening) and the technique shown in \cite{DLK13} (to suppress decoherence), the light storage efficiency and period could be significantly extended also inside a HCPCF, reaching and surpassing the values from free-space experiments \cite{CLW13}.

Finally, we turn to the creation of SLPs. Figure~\ref{fig:SL_LS_SLP}(c) shows the transmission of an incident probe pulse (black squares) through the HCPCF when driven under slow light and SLP conditions ($N_{\mathrm{avg}}$=250, $\tau_g$=100~ns).
As before, $\Omega_c^+$ was first adjusted (with $\Omega_c^-$=0) to fit the probe pulse well into the medium (orange circles).
Then the counterpropagating control (red dashed line) was applied as the probe pulse was inside the medium.
During the time when the medium is driven by the two counterpropagating control fields the transmission through the fiber is significantly suppressed (red diamonds). Switching off the backward control field again, retrieves the remaining coherence, i.e., a light pulse from the medium (see Fig.~\ref{fig:SL_LS_SLP}(c) inset). The retrieval efficiency is $\eta$=$2.8(6)~\%$ and occurs at times where there should not be anymore coherence left inside the medium for a continuously propagating pulse.
This is the typical signature of SLPs at moderate ODs when the probe pulse just fits inside the medium \cite{BZL03,LLP09}. 
The transmitted light detected when $\Omega_c^\pm(t)$$\neq$0 can be explained by the parts of the pulse near the edges of the medium leaking out due to diffusion \cite{BZL03,LLP09}. 
This is also confirmed by the numerical simulation (red solid line).
As the two-photon Doppler shifts $(k_p^\pm$-$k_c^\mp)v$ were smaller than $\Delta\omega_{\mathrm{EIT}}$, we applied a relative detuning $(\Delta_c^+$-$\Delta_c^-)$=$3.5\Gamma$$>$$\Delta\omega_{\mathrm{EIT}}$=2.9$\Gamma$ 
to avoid excitation of coherences suppressing the SLPs \cite{LLP09,WAL10b,PSC12}. 
For $\Delta_c^-$=$\Delta_c^+$ no pulse could be retrieved, as expected for a cold medium \cite{LLP09}. 
Due to the ground state frequency difference of $\Delta\omega_{21}$=$2\pi$$\times$6.835~GHz and the exact 1D alignment of all laser beams a phase mismatch is present which must be compensated by a two-photon detuning $\delta_{pm}$=$-\Delta\omega_{21}v_g/c$ \cite{ZAL06}. 
If $\delta_{pm}$$\gtrsim$$\Delta\omega_{\mathrm{EIT}}$ this leads to strong attenuation of the SLPs. This effect becomes negligible for media of large OD and correspondingly large group delays, however, it has to be considered in our experiment. Therefore, we set the detunings $\Delta_{p,c}^+$ such that an effective negative two-photon detuning within the EIT window width was realized (taking into account the ac Stark shifts by the control). Without an initial two-photon detuning, no SLPs could be observed.
Unlike for phase-matched conditions \cite{LLP09}, we obtained the largest retrieval efficiency for $\Omega_c^-$$\neq$$\Omega_c^+$. This is confirmed by numerical simulations for our parameters with comparable $\delta_{pm}$, $\Delta\omega_{\mathrm{EIT}}$, and pulse bandwidth $\Delta\omega_p$. The simulations show that the ratio $\Omega_c^-/\Omega_c^+$$\rightarrow$1 for obtaining SLPs as $\Delta\omega_p$, $\delta_{pm}$$\ll$$\Delta\omega_{\mathrm{EIT}}$. Our explanation for this is as follows: The phase mismatch is relevant only for excitation of the backward propagating field \cite{SM}. When not all frequencies of the probe pulse can be phase matched, the resulting suppression of the backward propagating field has to be compensated by a stronger coupling $\Omega_c^-$ to achieve an effective SLP. As $\Delta\omega_p$, $\delta_{pm}$$\ll$$\Delta\omega_{\mathrm{EIT}}$ good phase matching for all probe frequencies becomes possible and SLPs are formed again for balanced coupling.

\textit{Outlook.}---
In view of the goal of achieving efficient light storage and NLO interactions at the single-photon limit the following steps have to be taken. (i) The relative extinction ratio of probe and control beams must be improved by $\sim$20~dB to reach the interesting single-photon regime. This is technically feasible by improved polarization filtering
as shown for a similar filter \cite{PML12}. 
(ii) Incoherent (off-resonant) absorption must be suppressed to allow for larger ODs and hence larger light storage and SLP efficiencies. This can be achieved by using the D$_1$ instead of the D$_2$ line with a fewer number of excited states and the approx. 5-fold larger hyperfine splitting \cite{Steck2015}. 
(iii) Inhomogeneous broadening by the control itself must be reduced to maintain the condition $\Delta\omega_{\mathrm{EIT}}$$\gg$$\gamma_{21}$ also at larger ODs. Cooling the atoms \textit{inside} the 1D FORT, as, e.g.,\ demonstrated in \cite{PBH12,WDB99}, 
will reduce decoherence, inhomogeneous broadening, and heating-induced losses, resulting in larger probe transmission, ODs, and longer averaging times. Collisional thermalization in 3D will be possible at our present atomic densities of $10^{12}$~cm$^{-3}$ \cite{BHP14} 
resulting in temperatures well below the Doppler limit \cite{WDB99}. 

In conclusion, we demonstrated the creation of SLPs and narrowband light storage using EIT inside a HCPCF for the first time. Good agreement between numerical simulations and the experiments was found when considering the radially inhomogeneous laser beams and atomic density inside the fiber. The light storage efficiency was limited to around 23(5)~\% at a decoherence rate of $\gamma_{21}$=$2\pi$$\times$$64~$kHz dominated by transit relaxation decay. The minimum number of photons per pulse was $\sim$70. The OD per atom was OD$^*$=0.0037(6) for the transition used, in agreement with previous results.
We discussed several strategies for improving the experiment towards the goal of reaching linear and NLO interactions at the single-photon regime. Our work therefore paves the way towards a multitude of experiments requiring strong light-matter interactions in the field of quantum and nonlinear optics.

\begin{acknowledgments}
The authors thank
M. Szarafanowicz and Z. Zhou for technical assistance,
T. Walther for providing us with a high-bandwidth ultra-low noise current driver, and B.W. Shore, R. Walser, and M. Fleischhauer for fruitful discussions.
The research leading to these results has received funding from the Deutsche Forschungsgemeinschaft and the European Union Seventh Framework Programme (FP7/2007-2013) under grant agreement
n$^\circ$ PCIG09-GA-2011-289305.
\end{acknowledgments}

%

\newpage
\begin{center}
\fontsize{14}{16}\selectfont
\chapter{\textbf{Supplemental Material to: \\
Stationary light pulses and narrowband light storage in a laser-cooled ensemble loaded into a hollow-core fiber}}
\end{center}
\fontsize{10}{12}\selectfont
\begin{center}
\section{Theoretical Description}

\subsection{Hamiltonian}
\end{center}

\label{sec:Hamiltonian}
\begin{figure}[h!]
\includegraphics[width= 12cm]{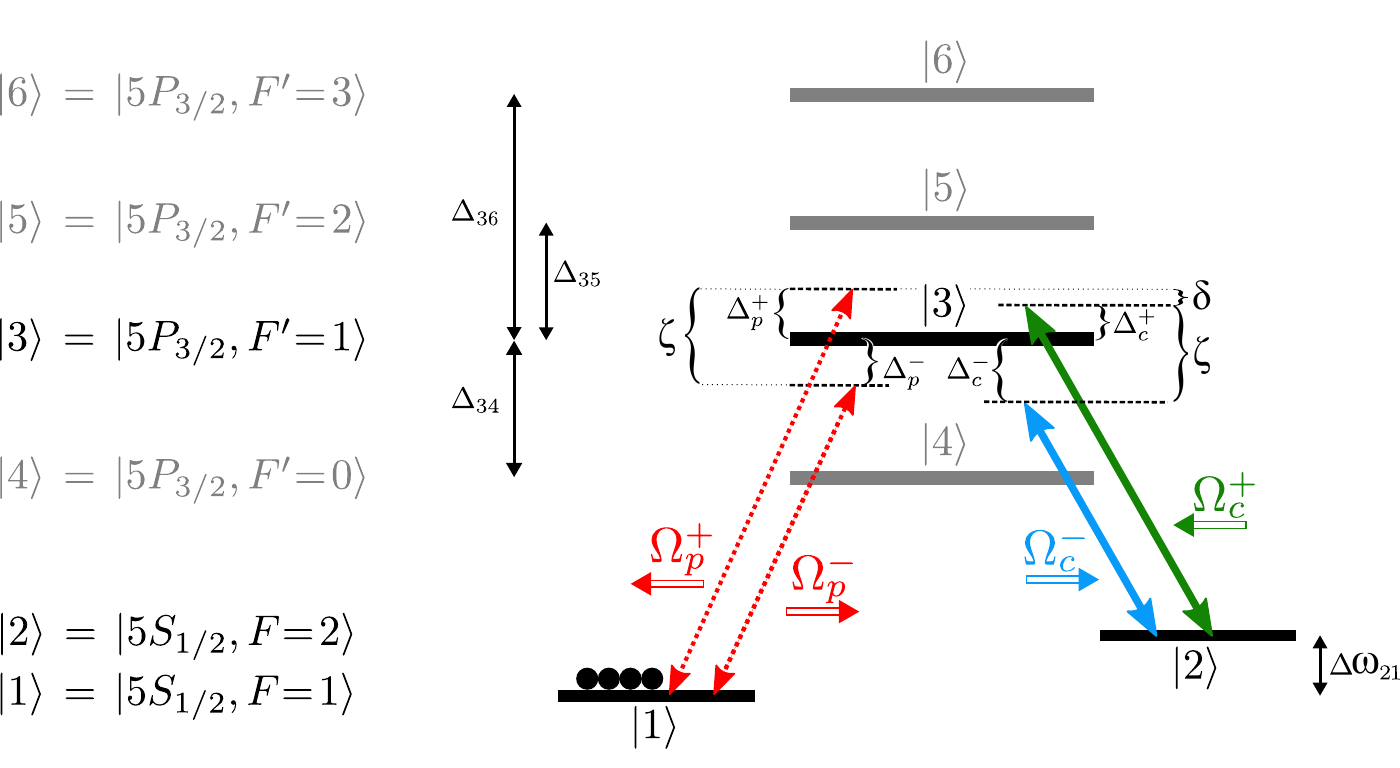}
\colorcaption{\label{fig:lambda_scheme}Level scheme of $^{87}$Rb used for the simulation of transmission and propagation.}
\end{figure}
We consider the level scheme corresponding to the D$_2$ line of $^{87}$Rb with degenerate Zeeman levels shown in Fig.~\ref{fig:lambda_scheme} with $\Delta_{3i}=\omega_i-\omega_3$ being the frequency difference between states $|i\rangle$ and $|3\rangle$. 
The system is driven by two strong counterpropagating control fields of Rabi frequencies $\Omega_c^\pm$ detuned from the transition $|5^2\mathrm{S}_{1/2},F$=$2\rangle\rightarrow|5^2\mathrm{P}_{3/2},F'$=$1\rangle$ by $\Delta_c^\pm=\omega_c^\pm -(\omega_3-\omega_2)$. A weak probe field $E_p^+$ of Rabi frequency $\Omega_p^+$ and detuned from the transition $|5^2\mathrm{S}_{1/2},F$=$1\rangle\rightarrow|5^2\mathrm{P}_{3/2},F'$=$1\rangle$ by $\Delta_p^+=\omega_{p}^{+}-(\omega_3-\omega_1)$ is sent into the medium propagating into the forward direction. Upon interaction with the two counterpropagating control fields, another probe field $E_p^-$ traveling into the backward direction is created in a four-wave mixing process with $\Delta_p^-=\Delta_p^+-\Delta_c^++\Delta_c^-$ due to energy conservation and Rabi frequency $\Omega_p^-$.
All 6 levels of the atomic structure are taken into account with level $|5^2\mathrm{P}_{3/2},F'$=$3\rangle$ being adiabatically eliminated.

The Hamiltonian reads
\begin{equation}
\hat{H}=\hbar(\omega_{p}^{+}-\omega_{c}^{+}-\Delta_{p}^{+}+\Delta_{c}^{+})|2\rangle\langle 2|+\hbar(\omega_{p}^{+}-\Delta_p^{+})|3\rangle\langle 3|+\hbar(\omega_{p}^{+}-\Delta_p^{+}+\Delta_{34})|4\rangle\langle 4|+\hbar(\omega_{p}^{+}-\Delta_p^{+}+\Delta_{35})|5\rangle\langle 5|+\hat{V},
\end{equation}
where $\hat{V}$ is the interaction part with the electromagnetic field,
\begin{align}
-\frac{2}{\hbar}\hat{V}=&e^{-\text{i}\omega_{p}^{+}t}\sum_{\mu=3}^{5}\left(\Omega_{p}^{+(1\mu)}e^{\text{i}kz}+\Omega_{p}^{-(1\mu)}e^{-\text{i}\zeta t-\text{i}kz}\right)|\mu\rangle\langle 1|\notag\\ &+e^{-\text{i}\omega_{c}^{+}t}\sum_{\mu=3}^{5}\left(\Omega_{c}^{+(2\mu)}e^{\text{i}kz}+\Omega_{c}^{-(2\mu)}e^{-\text{i}\zeta t-\text{i}kz}\right)|\mu\rangle\langle 2|+\text{h.c.},
\end{align}
with $\zeta\equiv\Delta_c^{-}-\Delta_c^{+}=\Delta_p^{-}-\Delta_p^{+}$, 
$\Omega_{c}^{\pm(24)}=0$ due to selection rules, and $k=k_{c}\approx k_p$ is the wave vector of the control field. $\Omega_{p}^{\pm(1\mu)}$ denote the probe beam Rabi frequencies of the fields traveling along the $\pm z$-direction. The superscripts $(1\mu)$ imply that the probe electric field $E_{p}$ is multiplied with the corresponding normalized relative hyperfine transition strength factors $\tilde{S}_{1\mu}=\sqrt{S_{1\mu}/S_{13}}$ \cite{Steck2015SM}
for the transition $|1\rangle\leftrightarrow|\mu\rangle$, normalized to the transition $|1\rangle\leftrightarrow|3\rangle$. Thus, we assume unpolarized excitation. 
The same applies for the control beam Rabi frequencies $\Omega_{c}^{\pm(2\mu)}$ with  $\tilde{S}_{2\mu}=\sqrt{S_{2\mu}/S_{23}}$ \cite{Steck2015SM} 
normalized to the transition $|2\rangle\leftrightarrow|3\rangle$. For simplicity we use below the notation $\Omega_{p}^{\pm(13)}\equiv\Omega_{p}^{\pm}$ as well as $\Omega_{c}^{\pm(23)}\equiv\Omega_{c}^{\pm}$ as the near-resonant transitions in our experiments are $|1\rangle\leftrightarrow|3\rangle\leftrightarrow|2\rangle$.
As state $|5^2\mathrm{P}_{3/2},F'$=$3\rangle$ is adiabatically eliminated, it appears in form of a Stark shift 
\begin{equation}
\Delta_S=\Delta_S^{+}+\Delta_S^{-} \;\;\;\; \textrm{with} \;\;\;\; \Delta_S^\pm = -\frac{|\Omega_c^{\pm(26)}|^{2}}{4(\Delta_{36}-\Delta_c^\pm)}.
\end{equation}
For convenience we neglect this Stark shift in the following, but it will be included later on by appropriate redefinition of the detunings $\Delta_c^{\pm}$. 

\subsection{Pulse propagation}
\label{sec:SLP}

The relevant Bloch equations \cite{WAL10bSM} of the 5-level system considering selection rules are then given by
\begin{align}
\frac{\partial}{\partial t}\rho_{21}=&-\left[\frac{\gamma_{21}}{2}-\text{i}\delta\right]\rho_{21}+\frac{\text{i}}{2}\left(\Omega_{c}^{+*}e^{-\text{i}kz}+\Omega_{c}^{-*}e^{\text{i}\zeta t+\text{i}kz}\right)\rho_{31}+\frac{\text{i}}{2}\left(\Omega_{c}^{+(25)*}e^{-\text{i}kz}+\Omega_{c}^{-(25)*}e^{\text{i}\zeta t+\text{i}kz}\right)\rho_{51},\label{eqn:MB1}\\
\frac{\partial}{\partial t}\rho_{31}=&-\left[\frac{\Gamma}{2}-\text{i}\Delta_p^{+}\right]\rho_{31}+\frac{\text{i}}{2}\left(\Omega_{p}^{+}e^{\text{i}kz}+\Omega_{p}^{-}e^{-\text{i}\zeta t-\text{i}kz}\right)+\frac{\text{i}}{2}\left(\Omega_{c}^{+}e^{\text{i}kz}+\Omega_{c}^{-}e^{-\text{i}\zeta t-\text{i}kz}\right)\rho_{21},\\
\frac{\partial}{\partial t}\rho_{41}=&-\left[\frac{\Gamma}{2}-\text{i}(\Delta_p^{+}-\Delta_{34})\right]\rho_{41}+\frac{i}{2}\left(\Omega_{p}^{+(14)}e^{\text{i}kz}+\Omega_{p}^{-(14)}e^{-\text{i}\zeta t-\text{i}kz}\right),\\
\frac{\partial}{\partial t}\rho_{51}=&-\left[\frac{\Gamma}{2}-\text{i}(\Delta_p^{+}-\Delta_{35})\right]\rho_{51}+\frac{\text{i}}{2}\left(\Omega_{p}^{+(15)}e^{\text{i}kz}+\Omega_{p}^{-(15)}e^{-\text{i}\zeta t-\text{i}kz}\right)\notag\\
&+\frac{\text{i}}{2}\left(\Omega_{c}^{+(25)}e^{ikz}+\Omega_{c}^{-(25)}e^{-\text{i}\zeta t-\text{i}kz}\right)\rho_{21}.\label{eqn:MB4}
\end{align}
Here, $\rho_{jk}$ are the matrix elements of the density matrix $\langle j|\hat{\rho}|k\rangle$ between the atomic energy states $|j\rangle$ and $|k\rangle$, $\gamma_{21}$ is the ground state decoherence rate, $\Gamma$ is the excited state decay rate, 
and $\delta=\Delta_p^{\pm}-\Delta_c^{\pm}$ is the two-photon detuning (according to sending the pulses into the medium with a forward control field).
We use $\rho_{11}^{(0)}=1$ and $\rho_{ij}^{(0)}=0$ for all $(i,j)\neq (1,1)$ as initial conditions, i.e., all population is in state $|1\rangle$.

Next, we expand the coherences into spatial Fourier components \cite{ZAL06SM,WAL10bSM}
\begin{align}
&\rho_{21}=\sum_{n=-\infty}^{+\infty}\rho_{21}^{(2n)}e^{\text{i}n\zeta t+\text{i}2nkz},\\
&\rho_{31}=\sum_{n=0}^{\infty}\rho_{31}^{(2n+1)}e^{\text{i}n\zeta t+\text{i}(2n+1)kz}+\sum_{n=-\infty}^{0}\rho_{31}^{(2n-1)}e^{\text{i}(n-1)\zeta t+\text{i}(2n-1)kz},\\
&\rho_{41}=\rho_{41}^{(+1)}e^{\text{i}kz}+\rho_{41}^{(-1)}e^{-\text{i}\zeta t-\text{i}kz},\\
&\rho_{51}=\sum_{n=0}^{\infty}\rho_{51}^{(2n+1)}e^{\text{i}n\zeta t+\text{i}(2n+1)kz}+\sum_{n=-\infty}^{0}\rho_{51}^{(2n-1)}e^{\text{i}(n-1)\zeta t+\text{i}(2n-1)kz},
\end{align}
and substitute these into the Bloch equations (\ref{eqn:MB1})-(\ref{eqn:MB4}). After substitution we obtain an infinite series of coherences spatially varying as $\mathrm{e}^{\pm \text{i} (2n+1)kz}$ and $\mathrm{e}^{\pm \text{i} 2nkz}$ with $n\geq0$. For practical purposes this series can be truncated for a suitable $n>n_{max}$ determined by the temperature of the medium or the relative detuning $|\zeta|$ \cite{WAL10bSM,PSC12SM}. While $n_{max}=0$ for a room-temperature medium, this approximation can also be done for cold atoms when the relative detunings of the counterpropagating control fields or the corresponding Doppler shifts $\pm2nkv_{atom}$ of the coherences $\rho_{21}^{\pm 2n}$ are much larger than the EIT transparency window width $\Delta\omega_{\mathrm{EIT}}$ \cite{WAL10bSM,PSC12SM}.
We therefore obtain for the Maxwell-Bloch equations describing the propagation dynamics
\begin{align}
\frac{\partial}{\partial t}\rho_{21}^{\pm(2n)}\ \ \ =&-\left[\frac{\gamma_{21}}{2}-\text{i}(\delta \mp \! \Delta_n^+)\right]\rho_{21}^{\pm(2n)}+\frac{\text{i}}{2}\Omega_{c}^{+*}\rho_{31}^{(\pm 2n+1)}+\frac{\text{i}}{2}\Omega_{c}^{-*}\rho_{31}^{(\pm 2n-1)}+\frac{\text{i}}{2}\tilde{S}_{25}\Omega_{c}^{+*}\rho_{51}^{(\pm 2n+1)}\notag\\
&+\frac{\text{i}}{2}\tilde{S}_{25}\Omega_{c}^{-*}\rho_{51}^{(\pm 2n-1)},\label{eqn:MB12} \\
\frac{\partial}{\partial t}\rho_{31}^{\pm(2n+1)}=&-\left[\frac{\Gamma}{2}-\text{i}\left(\Delta_p^{+}\mp\Delta_n^{\pm}\right)\right]\rho_{31}^{\pm(2n+1)}+\frac{\text{i}}{2}\Omega_{p}^{\pm}\delta_{n,0}+\frac{\text{i}}{2}\Omega_{c}^{\pm}\rho_{21}^{\pm(2n)}+\frac{\text{i}}{2}\Omega_{c}^{\mp}\rho_{21}^{\pm(2n+2)},\\
\frac{\partial}{\partial t}\rho_{41}^{(\pm1)}\ \ \ \ \ \!=&-\left[\frac{\Gamma}{2}-\text{i}(\Delta_p^{\pm}-\Delta_{34})\right]\rho_{41}^{(\pm1)}+\frac{\text{i}}{2}\tilde{S}_{14}\Omega_{p}^{\pm},\\
\frac{\partial}{\partial t}\rho_{51}^{\pm(2n+1)}=&-\left[\frac{\Gamma}{2}-\text{i}(\Delta_p^{+}-\Delta_{35} \mp \Delta_n^{\pm})\right]\rho_{51}^{\pm(2n+1)}+\frac{\text{i}}{2}\tilde{S}_{15}\Omega_{p}^{\pm}\delta_{n,0}+\frac{\text{i}}{2}\tilde{S}_{25}\Omega_{c}^{\pm}\rho_{21}^{\pm(2n)}\notag\\
&+\frac{\text{i}}{2}\tilde{S}_{25}\Omega_{c}^{\mp}\rho_{21}^{\pm(2n+2)},\\ 
\notag\\ 
\frac{1}{c}\frac{\partial}{\partial t}\Omega_{p}^{\pm}\pm\frac{\partial}{\partial z}\Omega_{p}^{\pm}=&\text{i}\frac{\Delta\omega_{21}}{c}\Omega_p^{\pm} + \text{i}\frac{\mathrm{OD}\:\Gamma}{2L}\left(\rho_{31}^{(\pm1)}+\tilde{S}_{14}\rho_{41}^{(\pm1)}+\tilde{S}_{15}\rho_{51}^{(\pm1)}\right),\label{eqn:MB16}
\end{align}
for $n\geq0$ with $\Delta_n^{+}=n\zeta, \Delta_n^{-}=(n+1)\zeta$, the Kronecker delta $\delta_{n,0}$, $\Delta_p^{-}=\Delta_p^{+}+\zeta$, OD being the optical depth for the transition $|1\rangle \leftrightarrow|3\rangle$ and $L$ being the medium length.

We also accounted for a phase mismatch $(k_p^+ - k_c^+)c=\Delta\omega_{21}=2\pi\times 6.835$~GHz due to the energy difference of the two ground states $|1\rangle$ and $|2\rangle$ and the perfect 1D geometry of the propagation inside the fiber \cite{ZAL06SM,WAL10bSM}. This phase mismatch is relevant for the excitation of the backward propagating field $\Omega_p^-$ ($\Delta K^- = | \vec{k}_p^{+} - \vec{k}_c^{+} + \vec{k}_c^{-} - \vec{k}_p^{-} | \approx 2(k_p^{+}-k_c^{+})$ with $k_{p,c}^{+}=|\vec{k}_{p,c}^{+}|$ and $\vec{k}_{p,c}^{+}\approx -\vec{k}_{p,c}^{-} $) only but not for $\Omega_p^+$ ($\Delta K^+ = k_p^+-k_c^++k_c^+-k_p^+=0$) when the probe pulses are sent into the medium with the forward control being present.

We note that the time-dependent Stark shift $\Delta_S$ is included in the detunings $\Delta_c^\pm$ and the Doppler shifts $\pm k v_{atom}$ are included in the detunings $\Delta_{p,c}^\pm$. 
An average over the thermal velocity distribution for given temperature $\Theta$ can therefore be taken \cite{WAL10bSM,PSC12SM}.
The ground state decoherence rate $\gamma_{21}=\gamma_{trd}+\gamma_{inh}$ is determined by transit relaxation decay $\gamma_{trd}$ (depending on the temperature $\Theta$ of the atoms inside the fiber) \cite{SNH96SM}, 
and an effective contribution $\gamma_{inh}$ due to the inhomogeneous broadening by the spatially varying control beams and atomic density (see Sec.~\ref{sec:EIT}). 
We neglect the contribution of the relative linewidth $\gamma_{lock}$ of the probe and control beams, since $\gamma_{lock}\ll\gamma_{21}$.
To further account for the radially inhomogeneous distributions of atomic density and Rabi frequencies, we also use effective Rabi frequencies $\Omega^{\mathit{eff}}=\beta\Omega$ determined from simulated transmission spectra (see Sec.\ref{sec:EIT}). We then numerically solve Eqns.~(\ref{eqn:MB12})-(\ref{eqn:MB16}) to simulate slow light, light storage and retrieval, and SLPs with $n_{max}$=$3$.

\subsection{EIT transmission spectra}
\label{sec:EIT}

We calculate transmission spectra $T(\Delta_p^{+})$ for the 6-level system, displayed in Fig.~\ref{fig:lambda_scheme} (state $|6\rangle$ is adiabatically eliminated) under EIT conditions with a single control field $\Omega_c^+$ as follows. We set $\Omega_{p,c}^{-}=\Delta_{p,c}^{-}\equiv 0$ and we substitute $\Omega_{p,c}^{+}=\Omega_{p,c}$ in the equations above. The Hamiltonian and the Bloch equations are therefore the same as in the previous section, however, with all couplings in the $-z$ direction set to 0.

For obtaining the stationary transmission, we derive the stationary solution of Eqns.~(\ref{eqn:MB1})-(\ref{eqn:MB4}), using the same initial conditions $\rho_{11}^{(0)}=1$ and $\rho_{ij}^{(0)}=0$ for all $(i,j)\neq (1,1)$ as before. From here we obtain the absorption coefficient for a \textit{homogeneous} medium as 
\begin{align}
\alpha(\Delta_p^{+})=A_{p4}\tilde{S}_{14}^2+\left[\sum_{k=3,5}A_{pk}\tilde{S}_{1k}^2+B A_{p3}A_{p5}|\tilde{S}_{15}-\tilde{S}_{25}|^{2}\Omega_{c}^2\right].\left[1+BA_{p3}|\Omega_{c}|^{2}+BA_{p5}|\tilde{S}_{25}\Omega_{c}|^{2}\right]^{-1},
\label{eqn:absorption}
\end{align}
where $A_{pj}=\left[\Gamma/2-\text{i}(\Delta_{p}^{+}-\xi_{j})\right]^{-1}$, $j=3,4,5$, and $\xi_{3}=0$, $\xi_{4}=\Delta_{34}$, $\xi_{5}=\Delta_{35}$ and\\ $4B=\left[\gamma_{21}/2-\text{i}(\Delta_{p}^{+}-\Delta_{c}^{+}-\Delta_S)\right]^{-1}$. We here have re-introduced the Stark shift $\Delta_S=\Delta_S^+$ due to level $|F^{\prime}=3\rangle$.

The transmission is then given by 
\begin{equation}
	T(\Delta_p^{+}) = \exp{\left[-\frac{\Gamma}{2}\mathrm{OD}\:\alpha(\Delta_p^{+})\right]},
\label{eqn:Thom}
\end{equation}
with $\mathrm{OD}=n_0\,\sigma_{atom}\,L$ where $n_0$ is the atomic density, $\sigma_{atom}$ is the absorption cross section, and $L$ is the length of the medium.

To include the radially inhomogeneous Rabi frequencies $\Omega(r)$ and atomic density $n(r)$ depending on the radial distance $r$ from the fiber axis, we obtain the transmission through the fiber with cylindrical symmetry as follows. The Rabi frequencies $\Omega_{p,c}(r)=\Omega_{p,c}(0) \exp{(-r^2/\sigma_{pc}^2)}$ with the measured mode field $^1\!\!/\!e^2$ radius of the intensity $\sigma_{pc}$ of the probe/control beams simply replaces the constant Rabi frequencies in Eqn.~(\ref{eqn:absorption}). The radial atomic density distribution is $n(r)=n_0 \exp{(-r^2/\sigma_a^2)}$, with the $^1\!\!/\!e^2$ radius $\sigma_a$ as determined by the temperature $\Theta$ of the atoms inside the FORT potential of known depth \cite{GWO00SM}.

The transmission through the \textit{inhomogeneous} medium inside the fiber is then given by
\begin{equation}
T(\Delta_p^{+},\Theta) = \exp\left[-\Gamma\frac{\sigma_{atom}}{\sigma_{pc}^2} L \int\!\!\!\int\!n(r) p(v_{atom}) \Omega_p(r)\alpha(r,\Delta_p^{+}) r \; dr dv_{atom}\right]
\label{eqn:Tinh}
\end{equation}
with the Maxwell-Boltzmann distribution $p(v_{atom})$ for the temperature $\Theta$.

The parameters $\beta$ and $\gamma_{inh}$ that include the effect of the inhomogeneities and are used for solving the Maxwell-Bloch equations (see Sec.~\ref{sec:SLP}) are determined by comparing the results of Eqns.~(\ref{eqn:Thom}) and (\ref{eqn:Tinh}) for the homogeneous/inhomogeneous cases with $\beta$ and $\gamma_{inh}$ included in the homogeneous transmission function. We then adjust $\beta$ and $\gamma_{inh}$ for all other parameters being identical, until the transmission of both spectra is the same within $<$1~\%. As $\beta$ and $\gamma_{inh}$ depend on the temperature $\Theta$ and the control Rabi frequencies, they are determined for each of them individually in the range of our experimental parameters.

\begin{center}
\section{Experimental Details}

\end{center}

\textit{Fiber loading procedure} -- 
We first loaded about $N_0~=~10^7$ rubidium atoms into a standard vapor cell magneto-optical trap (MOT) with rectangular coil geometry \cite{PWB12SM}.
After a loading period of 1~s we transferred the atom cloud down towards the tip of a vertically oriented hollow-core photonic crystal fiber (HCPCF, HC-800-02, NKT Photonics), located $\sim$$5.5$~mm away from the center of the MOT, by shifting the magnetic zero point of the MOT with an offset magnetic field. Simultaneously, we compressed the cloud by ramping up the current in the quadrupole coils of the MOT to achieve a gradient of 15~G/cm. 
To avoid density-limiting light-assisted collisions near the HCPCF, we used the so-called \textit{dark spot} technique \cite{KDJ93SM} to create a \textit{dark funnel} for the atoms \cite{BHP14SM}. 
While the atom cloud was held above the fiber tip they could fall into a near Gaussian-shaped red-detuned far-off-resonant trap (FORT) \cite{GWO00SM} located inside the HCPCF. The FORT was realized by coupling radiation at a wavelength of 855~nm and a power of 270~mW (corresponding to a trap depth of $\sim$5~mK) into the HCPCF. The FORT therefore prevented collisions of the laser-cooled rubidium atoms with the room-temperature fiber wall, allowing for guiding and a tight confinement of the atoms. With this setup, we were able to load up to 2.5~\% of the atoms into the HCPCF, resulting in an OD of up to 1000 \cite{BHP14SM}. 
The total number of atoms $N_{atom}$ loaded into the fiber could be controlled by the power of the repumper beam tuned to the transition $|1\rangle\rightarrow|5\rangle$ which forms the dark funnel. 
The loading process of the HCPCF was repeated every 1.3~s. For further experimental details on the loading procedure we refer the reader to \cite{BHP14SM}.
  \\

\textit{Detection of the atoms inside the fiber} -- 
To guarantee interaction with only the atoms inside the HCPCF during any of the here presented experiments, we switched off the magnetic field of the MOT 14~ms earlier and pumped the region above the fiber continuously with the \textit{repumper} beam (tuned to the transition $|1\rangle\rightarrow|5\rangle$) with a power of 600~$\mu$W and a diameter of 2~mm ($1/e^2$) once the fiber was loaded. The atoms inside the HCPCF were optically pumped into state $|F$=1$\rangle$ by the control laser.
  \\

\textit{Fast FORT modulation} -- 
The FORT was modulated by an acousto-optic modulator (AOM) placed between the diode laser and the tapered amplifier of our MOPA system \cite{BHP14SM}. Although the 90--20~\% fall time was 170~ns, the rf power of the AOM driver oscillator would decay with a time constant of 260~ns (1/e) thereafter leading to non-negligible ac Stark shifts during the measurements within the short measurement periods of $\tau_{meas}\leq 3~\mu$s. This effect was qualitatively observed for different commercial and home-built drivers. To overcome this problem we used an additional fast absorptive switch (MiniCircuits, ZYSWA-2-50DR) between oscillator and amplifier of the AOM driver. The resulting 90--10~\% fall- and risetime was then 95~ns with a suppression of $\geq 14$~dB of the FORT during the measurements. This is sufficient to reduce the trap depth to $<225~\mu$K, which is below the temperature of the atoms, and to an ac Stark shift of less than $0.2~\Gamma$ of the transition $|2\rangle\rightarrow|3\rangle$.
  \\

\textit{Magnetic shielding} -- 
Although the central part of the HCPCF was shielded from magnetic stray fields by a layer of $\mu$-metal, the regions within around 2~cm from the fiber tips were not shielded. We therefore observed an effect of the (decaying) quadrupole field of the MOT on the light storage efficiency by showing a beating of the retrieval efficiency \cite{PCW09SM,GMD15SM}. This effect could however be compensated by applying a suitable 3D magnetic offset field following the procedure in \cite{PCW09SM}. This offset field had to be adjusted slightly day by day.
  \\
	
\textit{Light Polarization} -- 
Probe and control laser beams were launched with linear orthogonal polarizations into the HCPCF. Due to the typical birefringence of HCPCFs \cite{SMU05SM} and non-perfect input coupling \cite{CLV04SM} the light fields were elliptically polarized inside the fiber (degree of linear polarization $\sim$$90~\%$ after the fiber). For the chosen coupling scheme, however, EIT conditions for all possible Zeeman transitions are achieved.
  \\

\textit{Probe detection} -- 
In order to separate the weak probe field (3.5~pW $\leq P_{probe}\leq700$~pW) from the much stronger and exactly collinear forward control field (50~nW $\leq P_{control}\leq$ 1.7~$\mu$W) after the HCPCF for detection with a photon counter (PerkinElmer, SPCM AQRH-12), we proceeded as follows. 
First, the light exiting the fiber was passing a polarization beam splitter. Then the light was spatially filtered by a single-mode fiber \cite{BHP11SM}. Finally we used a combination of a monolithic etalon, polarization filter and a second spatial filter \cite{PML12SM} by coupling the light into a single-mode fiber leading to the photon counter. This led to a detection efficiency of 10~\% for the probe and a relative extinction ratio of 64~dB for the control beam.
  \\


\begin{thebibliography}{51}%
\makeatletter
\providecommand \@ifxundefined [1]{%
 \@ifx{#1\undefined}
}%
\providecommand \@ifnum [1]{%
 \ifnum #1\expandafter \@firstoftwo
 \else \expandafter \@secondoftwo
 \fi
}%
\providecommand \@ifx [1]{%
 \ifx #1\expandafter \@firstoftwo
 \else \expandafter \@secondoftwo
 \fi
}%
\providecommand \natexlab [1]{#1}%
\providecommand \enquote  [1]{``#1''}%
\providecommand \bibnamefont  [1]{#1}%
\providecommand \bibfnamefont [1]{#1}%
\providecommand \citenamefont [1]{#1}%
\providecommand \href@noop [0]{\@secondoftwo}%
\providecommand \href [0]{\begingroup \@sanitize@url \@href}%
\providecommand \@href[1]{\@@startlink{#1}\@@href}%
\providecommand \@@href[1]{\endgroup#1\@@endlink}%
\providecommand \@sanitize@url [0]{\catcode `\\12\catcode `\$12\catcode
  `\&12\catcode `\#12\catcode `\^12\catcode `\_12\catcode `\%12\relax}%
\providecommand \@@startlink[1]{}%
\providecommand \@@endlink[0]{}%
\providecommand \url  [0]{\begingroup\@sanitize@url \@url }%
\providecommand \@url [1]{\endgroup\@href {#1}{\urlprefix }}%
\providecommand \urlprefix  [0]{URL }%
\providecommand \Eprint [0]{\href }%
\providecommand \doibase [0]{http://dx.doi.org/}%
\providecommand \selectlanguage [0]{\@gobble}%
\providecommand \bibinfo  [0]{\@secondoftwo}%
\providecommand \bibfield  [0]{\@secondoftwo}%
\providecommand \translation [1]{[#1]}%
\providecommand \BibitemOpen [0]{}%
\providecommand \bibitemStop [0]{}%
\providecommand \bibitemNoStop [0]{.\EOS\space}%
\providecommand \EOS [0]{\spacefactor3000\relax}%
\providecommand \BibitemShut  [1]{\csname bibitem#1\endcsname}%
\let\auto@bib@innerbib\@empty
\bibitem [{\citenamefont {Gorshkov}\ \emph {et~al.}(2007)\citenamefont
  {Gorshkov}, \citenamefont {Andr{\'{e}}}, \citenamefont {Lukin},and\
  \citenamefont {S{\o}rensen}}]{GAL07b}%
  \BibitemOpen
  \bibfield  {author} {\bibinfo {author} {\bibfnamefont {A.~V.}\ \bibnamefont
  {Gorshkov}}, \bibinfo {author} {\bibfnamefont {A.}~\bibnamefont
  {Andr{\'{e}}}}, \bibinfo {author} {\bibfnamefont {M.~D.}\ \bibnamefont
  {Lukin}}, and\ \bibinfo {author} {\bibfnamefont {A.~S.}\ \bibnamefont
  {S{\o}rensen}},\ }\href {\doibase 10.1103/PhysRevA.76.033805} {\bibfield
  {journal} {\bibinfo  {journal} {Physical Review A}\ }\textbf {\bibinfo
  {volume} {76}},\ \bibinfo {pages} {033805} (\bibinfo {year}
  {2007})}\BibitemShut {NoStop}%
\bibitem [{\citenamefont {Fleischhauer}\ \emph {et~al.}(2005)\citenamefont
  {Fleischhauer}, \citenamefont {Imamoglu},and\ \citenamefont
  {Marangos}}]{FIM05}%
  \BibitemOpen
  \bibfield  {author} {\bibinfo {author} {\bibfnamefont {M.}~\bibnamefont
  {Fleischhauer}}, \bibinfo {author} {\bibfnamefont {A.}~\bibnamefont
  {Imamoglu}}, and\ \bibinfo {author} {\bibfnamefont {J.~P.}\ \bibnamefont
  {Marangos}},\ }\href {\doibase 10.1103/RevModPhys.77.633} {\bibfield
  {journal} {\bibinfo  {journal} {Reviews of Modern Physics}\ }\textbf
  {\bibinfo {volume} {77}},\ \bibinfo {pages} {633} (\bibinfo {year}
  {2005})}\BibitemShut {NoStop}%
\bibitem [{\citenamefont {Chen}\ \emph {et~al.}(2013)\citenamefont {Chen},
  \citenamefont {Lee}, \citenamefont {Wang}, \citenamefont {Du}, \citenamefont
  {Chen}, \citenamefont {Chen},\ and\ \citenamefont {Yu}}]{CLW13}%
  \BibitemOpen
  \bibfield  {author} {\bibinfo {author} {\bibfnamefont {Y.-H.}\ \bibnamefont
  {Chen}}, \bibinfo {author} {\bibfnamefont {M.-J.}\ \bibnamefont {Lee}},
  \bibinfo {author} {\bibfnamefont {I.-C.}\ \bibnamefont {Wang}}, \bibinfo
  {author} {\bibfnamefont {S.}~\bibnamefont {Du}}, \bibinfo {author}
  {\bibfnamefont {Y.-F.}\ \bibnamefont {Chen}}, \bibinfo {author}
  {\bibfnamefont {Y.-C.}\ \bibnamefont {Chen}}, \ and\ \bibinfo {author}
  {\bibfnamefont {I.~A.}\ \bibnamefont {Yu}},\ }\href {\doibase
  10.1103/PhysRevLett.110.083601} {\bibfield  {journal} {\bibinfo  {journal}
  {Physical Review Letters}\ }\textbf {\bibinfo {volume} {110}},\ \bibinfo
  {pages} {083601} (\bibinfo {year} {2013})}\BibitemShut {NoStop}%
\bibitem [{\citenamefont {Schraft}\ \emph {et~al.}(2016)\citenamefont
  {Schraft}, \citenamefont {Hain}, \citenamefont {Lorenz},\ and\ \citenamefont
  {Halfmann}}]{SHL16}%
  \BibitemOpen
  \bibfield  {author} {\bibinfo {author} {\bibfnamefont {D.}~\bibnamefont
  {Schraft}}, \bibinfo {author} {\bibfnamefont {M.}~\bibnamefont {Hain}},
  \bibinfo {author} {\bibfnamefont {N.}~\bibnamefont {Lorenz}}, \ and\ \bibinfo
  {author} {\bibfnamefont {T.}~\bibnamefont {Halfmann}},\ }\href {\doibase
  10.1103/PhysRevLett.116.073602} {\bibfield  {journal} {\bibinfo  {journal}
  {Physical Review Letters}\ }\textbf {\bibinfo {volume} {116}},\ \bibinfo
  {pages} {073602} (\bibinfo {year} {2016})}\BibitemShut {NoStop}%
\bibitem [{\citenamefont {H{\'{e}}tet}\ \emph {et~al.}(2008)\citenamefont
  {H{\'{e}}tet}, \citenamefont {Longdell}, \citenamefont {Alexander},
  \citenamefont {Lam},\ and\ \citenamefont {Sellars}}]{HLA08}%
  \BibitemOpen
  \bibfield  {author} {\bibinfo {author} {\bibfnamefont {G.}~\bibnamefont
  {H{\'{e}}tet}}, \bibinfo {author} {\bibfnamefont {J.~J.}\ \bibnamefont
  {Longdell}}, \bibinfo {author} {\bibfnamefont {A.~L.}\ \bibnamefont
  {Alexander}}, \bibinfo {author} {\bibfnamefont {P.~K.}\ \bibnamefont {Lam}},
  \ and\ \bibinfo {author} {\bibfnamefont {M.~J.}\ \bibnamefont {Sellars}},\
  }\href {\doibase 10.1103/PhysRevLett.100.023601} {\bibfield  {journal}
  {\bibinfo  {journal} {Physical Review Letters}\ }\textbf {\bibinfo {volume}
  {100}},\ \bibinfo {pages} {023601} (\bibinfo {year} {2008})}\BibitemShut {NoStop}%
\bibitem [{\citenamefont {Cho}\ \emph {et~al.}(2016)\citenamefont {Cho},
  \citenamefont {Campbell}, \citenamefont {Everett}, \citenamefont {Bernu},
  \citenamefont {Higginbottom}, \citenamefont {Cao}, \citenamefont {Geng},
  \citenamefont {Robins}, \citenamefont {Lam},\ and\ \citenamefont
  {Buchler}}]{CCE16}%
  \BibitemOpen
  \bibfield  {author} {\bibinfo {author} {\bibfnamefont {Y.-W.}\ \bibnamefont
  {Cho}}, \bibinfo {author} {\bibfnamefont {G.~T.}\ \bibnamefont {Campbell}},
  \bibinfo {author} {\bibfnamefont {J.~L.}\ \bibnamefont {Everett}}, \bibinfo
  {author} {\bibfnamefont {J.}~\bibnamefont {Bernu}}, \bibinfo {author}
  {\bibfnamefont {D.~B.}\ \bibnamefont {Higginbottom}}, \bibinfo {author}
  {\bibfnamefont {M.~T.}\ \bibnamefont {Cao}}, \bibinfo {author} {\bibfnamefont
  {J.}~\bibnamefont {Geng}}, \bibinfo {author} {\bibfnamefont {N.~P.}\
  \bibnamefont {Robins}}, \bibinfo {author} {\bibfnamefont {P.~K.}\
  \bibnamefont {Lam}}, \ and\ \bibinfo {author} {\bibfnamefont {B.~C.}\
  \bibnamefont {Buchler}},\ }\href {\doibase 10.1364/OPTICA.3.000100}
  {\bibfield  {journal} {\bibinfo  {journal} {Optica}\ }\textbf {\bibinfo
  {volume} {3}},\ \bibinfo {pages} {100} (\bibinfo {year} {2016})}\BibitemShut {NoStop}%
\bibitem [{\citenamefont {Chang}\ \emph {et~al.}(2014)\citenamefont {Chang},
  \citenamefont {Vuleti{\'{c}}},and\ \citenamefont {Lukin}}]{CVL14}%
  \BibitemOpen
  \bibfield  {author} {\bibinfo {author} {\bibfnamefont {D.~E.}\ \bibnamefont
  {Chang}}, \bibinfo {author} {\bibfnamefont {V.}~\bibnamefont
  {Vuleti{\'{c}}}}, and\ \bibinfo {author} {\bibfnamefont {M.~D.}\
  \bibnamefont {Lukin}},\ }\href {\doibase 10.1038/nphoton.2014.192} {\bibfield
   {journal} {\bibinfo  {journal} {Nature Photonics}\ }\textbf {\bibinfo
  {volume} {8}},\ \bibinfo {pages} {685} (\bibinfo {year} {2014})}\BibitemShut
  {NoStop}%
\bibitem [{\citenamefont {Kimble}(2008)}]{Kimble2008}%
  \BibitemOpen
  \bibfield  {author} {\bibinfo {author} {\bibfnamefont {H.~J.}\ \bibnamefont
  {Kimble}},\ }\href {\doibase 10.1038/nature07127} {\bibfield  {journal}
  {\bibinfo  {journal} {Nature}\ }\textbf {\bibinfo {volume} {453}},\ \bibinfo
  {pages} {1023} (\bibinfo {year} {2008})}\BibitemShut {NoStop}%
\bibitem [{\citenamefont {Chang}\ \emph {et~al.}(2008)\citenamefont {Chang},
  \citenamefont {Gritsev}, \citenamefont {Morigi}, \citenamefont
  {Vuleti{\'{c}}}, \citenamefont {Lukin},and\ \citenamefont
  {Demler}}]{CGM08}%
  \BibitemOpen
  \bibfield  {author} {\bibinfo {author} {\bibfnamefont {D.~E.}\ \bibnamefont
  {Chang}}, \bibinfo {author} {\bibfnamefont {V.}~\bibnamefont {Gritsev}},
  \bibinfo {author} {\bibfnamefont {G.}~\bibnamefont {Morigi}}, \bibinfo
  {author} {\bibfnamefont {V.}~\bibnamefont {Vuleti{\'{c}}}}, \bibinfo {author}
  {\bibfnamefont {M.~D.}\ \bibnamefont {Lukin}}, and\ \bibinfo {author}
  {\bibfnamefont {E.~A.}\ \bibnamefont {Demler}},\ }\href {\doibase
  10.1038/nphys1074} {\bibfield  {journal} {\bibinfo  {journal} {Nature
  Physics}\ }\textbf {\bibinfo {volume} {4}},\ \bibinfo {pages} {884} (\bibinfo
  {year} {2008})}\BibitemShut {NoStop}%
\bibitem [{\citenamefont {Kiffner} and\ \citenamefont
  {Hartmann}(2010)}]{KH10b}%
  \BibitemOpen
  \bibfield  {author} {\bibinfo {author} {\bibfnamefont {M.}~\bibnamefont
  {Kiffner}} and\ \bibinfo {author} {\bibfnamefont {M.~J.}\ \bibnamefont
  {Hartmann}},\ }\href {\doibase 10.1103/PhysRevA.81.021806} {\bibfield
  {journal} {\bibinfo  {journal} {Physical Review A}\ }\textbf {\bibinfo
  {volume} {81}},\ \bibinfo {pages} {021806} (\bibinfo {year}
  {2010})}\BibitemShut {NoStop}%
\bibitem [{\citenamefont {Angelakis}\ \emph {et~al.}(2011)\citenamefont
  {Angelakis}, \citenamefont {Huo}, \citenamefont {Kyoseva},and\
  \citenamefont {Kwek}}]{AHK11}%
  \BibitemOpen
  \bibfield  {author} {\bibinfo {author} {\bibfnamefont {D.~G.}\ \bibnamefont
  {Angelakis}}, \bibinfo {author} {\bibfnamefont {M.}\ \bibnamefont {Huo}},
  \bibinfo {author} {\bibfnamefont {E.}~\bibnamefont {Kyoseva}}, and\
  \bibinfo {author} {\bibfnamefont {L.~C.}~\bibnamefont {Kwek}},\ }\href {\doibase
  10.1103/PhysRevLett.106.153601} {\bibfield  {journal} {\bibinfo  {journal}
  {Physical Review Letters}\ }\textbf {\bibinfo {volume} {106}},\ \bibinfo
  {pages} {153601} (\bibinfo {year} {2011})}\BibitemShut {NoStop}%
\bibitem [{\citenamefont {Hafezi}\ \emph {et~al.}(2012)\citenamefont {Hafezi},
  \citenamefont {Chang}, \citenamefont {Gritsev}, \citenamefont {Demler},and\
  \citenamefont {Lukin}}]{HCG12}%
  \BibitemOpen
  \bibfield  {author} {\bibinfo {author} {\bibfnamefont {M.}~\bibnamefont
  {Hafezi}}, \bibinfo {author} {\bibfnamefont {D.~E.}\ \bibnamefont {Chang}},
  \bibinfo {author} {\bibfnamefont {V.}~\bibnamefont {Gritsev}}, \bibinfo
  {author} {\bibfnamefont {E.}~\bibnamefont {Demler}}, and\ \bibinfo {author}
  {\bibfnamefont {M.~D.}\ \bibnamefont {Lukin}},\ }\href {\doibase
  10.1103/PhysRevA.85.013822} {\bibfield  {journal} {\bibinfo  {journal}
  {Physical Review A}\ }\textbf {\bibinfo {volume} {85}},\ \bibinfo {pages}
  {013822} (\bibinfo {year} {2012})}\BibitemShut {NoStop}%
\bibitem [{\citenamefont {Angelakis}\ \emph {et~al.}(2013)\citenamefont
  {Angelakis}, \citenamefont {Huo}, \citenamefont {Chang}, \citenamefont
  {Kwek},and\ \citenamefont {Korepin}}]{AHC13}%
  \BibitemOpen
  \bibfield  {author} {\bibinfo {author} {\bibfnamefont {D.~G.}\ \bibnamefont
  {Angelakis}}, \bibinfo {author} {\bibfnamefont {M.-X.}\ \bibnamefont {Huo}},
  \bibinfo {author} {\bibfnamefont {D.~E.}\ \bibnamefont {Chang}}, \bibinfo
  {author} {\bibfnamefont {L.~C.}\ \bibnamefont {Kwek}}, and\ \bibinfo
  {author} {\bibfnamefont {V.}~\bibnamefont {Korepin}},\ }\href {\doibase
  10.1103/PhysRevLett.110.100502} {\bibfield  {journal} {\bibinfo  {journal}
  {Physical Review Letters}\ }\textbf {\bibinfo {volume} {110}},\ \bibinfo
  {pages} {100502} (\bibinfo {year} {2013})}\BibitemShut {NoStop}%
\bibitem [{\citenamefont {Peyronel}\ \emph
  {et~al.}(2012{\natexlab{a}})\citenamefont {Peyronel}, \citenamefont
  {Firstenberg}, \citenamefont {Liang}, \citenamefont {Hofferberth},
  \citenamefont {Gorshkov}, \citenamefont {Pohl}, \citenamefont {Lukin},and\
  \citenamefont {Vuleti{\'{c}}}}]{PFL12}%
  \BibitemOpen
  \bibfield  {author} {\bibinfo {author} {\bibfnamefont {T.}~\bibnamefont
  {Peyronel}}, \bibinfo {author} {\bibfnamefont {O.}~\bibnamefont
  {Firstenberg}}, \bibinfo {author} {\bibfnamefont {Q.-Y.}\ \bibnamefont
  {Liang}}, \bibinfo {author} {\bibfnamefont {S.}~\bibnamefont {Hofferberth}},
  \bibinfo {author} {\bibfnamefont {A.~V.}\ \bibnamefont {Gorshkov}}, \bibinfo
  {author} {\bibfnamefont {T.}~\bibnamefont {Pohl}}, \bibinfo {author}
  {\bibfnamefont {M.~D.}\ \bibnamefont {Lukin}}, and\ \bibinfo {author}
  {\bibfnamefont {V.}~\bibnamefont {Vuleti{\'{c}}}},\ }\href {\doibase
  10.1038/nature11361} {\bibfield  {journal} {\bibinfo  {journal} {Nature}\
  }\textbf {\bibinfo {volume} {488}},\ \bibinfo {pages} {57} (\bibinfo {year}
  {2012}{\natexlab{a}})}\BibitemShut {NoStop}%
\bibitem [{\citenamefont {Dudin} and\ \citenamefont {Kuzmich}(2012)}]{DK12}%
  \BibitemOpen
  \bibfield  {author} {\bibinfo {author} {\bibfnamefont {Y.~O.}\ \bibnamefont
  {Dudin}} and\ \bibinfo {author} {\bibfnamefont {A.}~\bibnamefont
  {Kuzmich}},\ }\href {\doibase 10.1126/science.1217901} {\bibfield  {journal}
  {\bibinfo  {journal} {Science}\ }\textbf {\bibinfo {volume}
  {336}},\ \bibinfo {pages} {887} (\bibinfo {year} {2012})}\BibitemShut
  {NoStop}%
\bibitem [{\citenamefont {Firstenberg}\ \emph {et~al.}(2013)\citenamefont
  {Firstenberg}, \citenamefont {Peyronel}, \citenamefont {Liang}, \citenamefont
  {Gorshkov}, \citenamefont {Lukin},and\ \citenamefont
  {Vuleti{\'{c}}}}]{FPL13}%
  \BibitemOpen
  \bibfield  {author} {\bibinfo {author} {\bibfnamefont {O.}~\bibnamefont
  {Firstenberg}}, \bibinfo {author} {\bibfnamefont {T.}~\bibnamefont
  {Peyronel}}, \bibinfo {author} {\bibfnamefont {Q.-Y.}\ \bibnamefont {Liang}},
  \bibinfo {author} {\bibfnamefont {A.~V.}\ \bibnamefont {Gorshkov}}, \bibinfo
  {author} {\bibfnamefont {M.~D.}\ \bibnamefont {Lukin}}, and\ \bibinfo
  {author} {\bibfnamefont {V.}~\bibnamefont {Vuleti{\'{c}}}},\ }\href {\doibase
  10.1038/nature12512} {\bibfield  {journal} {\bibinfo  {journal} {Nature}\
  }\textbf {\bibinfo {volume} {502}},\ \bibinfo {pages} {71} (\bibinfo {year}
  {2013})}\BibitemShut {NoStop}%
\bibitem [{\citenamefont {Pritchard}\ \emph {et~al.}(2013)\citenamefont
  {Pritchard}, \citenamefont {Weatherill},and\ \citenamefont
  {Adams}}]{PWA13}%
  \BibitemOpen
  \bibfield  {author} {\bibinfo {author} {\bibfnamefont {J.}~\bibnamefont
  {Pritchard}}, \bibinfo {author} {\bibfnamefont {K.}~\bibnamefont
  {Weatherill}}, and\ \bibinfo {author} {\bibfnamefont {C.}~\bibnamefont
  {Adams}},\ }\href {\doibase 10.1142/9789814440400_0008} {\bibfield
  {journal} {\bibinfo  {journal} {Annual Review of Cold Atoms and Molecules}\ }\textbf
  {\bibinfo {volume} {1}},\ \bibinfo {pages} {301} (\bibinfo {year} {2013})}\BibitemShut
  {NoStop}%
\bibitem [{\citenamefont {Birnbaum}\ \emph {et~al.}(2005)\citenamefont
  {Birnbaum}, \citenamefont {Boca}, \citenamefont {Miller}, \citenamefont
  {Boozer}, \citenamefont {Northup},\ and\ \citenamefont
  {Kimble}}]{Birnbaum2005a}%
  \BibitemOpen
  \bibfield  {author} {\bibinfo {author} {\bibfnamefont {K.}~\bibnamefont
  {Birnbaum}}, \bibinfo {author} {\bibfnamefont {A.}~\bibnamefont {Boca}},
  \bibinfo {author} {\bibfnamefont {R.}~\bibnamefont {Miller}}, \bibinfo
  {author} {\bibfnamefont {A.}~\bibnamefont {Boozer}}, \bibinfo {author}
  {\bibfnamefont {T.}~\bibnamefont {Northup}}, \ and\ \bibinfo {author}
  {\bibfnamefont {H.}~\bibnamefont {Kimble}},\ }\href {\doibase
  10.1038/nature03804} {\bibfield  {journal} {\bibinfo  {journal} {Nature}\
  }\textbf {\bibinfo {volume} {436}},\ \bibinfo {pages} {87} (\bibinfo {year}
  {2005})}\BibitemShut {NoStop}%
\bibitem [{\citenamefont {Dayan}\ \emph {et~al.}(2008)\citenamefont {Dayan},
  \citenamefont {Parkins}, \citenamefont {Aoki}, \citenamefont {Ostby},
  \citenamefont {Vahala},\ and\ \citenamefont {Kimble}}]{Dayan2008}%
  \BibitemOpen
  \bibfield  {author} {\bibinfo {author} {\bibfnamefont {B.}~\bibnamefont
  {Dayan}}, \bibinfo {author} {\bibfnamefont {A.}~\bibnamefont {Parkins}},
  \bibinfo {author} {\bibfnamefont {T.}~\bibnamefont {Aoki}}, \bibinfo {author}
  {\bibfnamefont {E.}~\bibnamefont {Ostby}}, \bibinfo {author} {\bibfnamefont
  {K.}~\bibnamefont {Vahala}}, \ and\ \bibinfo {author} {\bibfnamefont
  {H.}~\bibnamefont {Kimble}},\ }\href {\doibase 10.1126/science.1152261}
  {\bibfield  {journal} {\bibinfo  {journal} {Science}\ }\textbf
  {\bibinfo {volume} {319}},\ \bibinfo {pages} {1062} (\bibinfo {year}
  {2008})}\BibitemShut {NoStop}%
\bibitem [{\citenamefont {Bajcsy}\ \emph {et~al.}(2003)\citenamefont {Bajcsy},
  \citenamefont {Zibrov},and\ \citenamefont {Lukin}}]{BZL03}%
  \BibitemOpen
  \bibfield  {author} {\bibinfo {author} {\bibfnamefont {M.}~\bibnamefont
  {Bajcsy}}, \bibinfo {author} {\bibfnamefont {A.~S.}\ \bibnamefont {Zibrov}},
  and\ \bibinfo {author} {\bibfnamefont {M.~D.}\ \bibnamefont {Lukin}},\
  }\href {\doibase 10.1038/nature02176} {\bibfield  {journal} {\bibinfo
  {journal} {Nature}\ }\textbf {\bibinfo {volume} {426}},\ \bibinfo {pages}
  {638} (\bibinfo {year} {2003})}\BibitemShut {NoStop}%
\bibitem [{\citenamefont {Goban}\ \emph {et~al.}(2014)\citenamefont {Goban},
  \citenamefont {Hung}, \citenamefont {Yu}, \citenamefont {Hood}, \citenamefont
  {Muniz}, \citenamefont {Lee}, \citenamefont {Martin}, \citenamefont
  {McClung}, \citenamefont {Choi}, \citenamefont {Chang}, \citenamefont
  {Painter},and\ \citenamefont {Kimble}}]{GHY14}%
  \BibitemOpen
  \bibfield  {author} {\bibinfo {author} {\bibfnamefont {A.}~\bibnamefont
  {Goban}}, \bibinfo {author} {\bibfnamefont {C.-L.}\ \bibnamefont {Hung}},
  \bibinfo {author} {\bibfnamefont {S.-P.}\ \bibnamefont {Yu}}, \bibinfo
  {author} {\bibfnamefont {J.}~\bibnamefont {Hood}}, \bibinfo {author}
  {\bibfnamefont {J.}~\bibnamefont {Muniz}}, \bibinfo {author} {\bibfnamefont
  {J.}~\bibnamefont {Lee}}, \bibinfo {author} {\bibfnamefont {M.}~\bibnamefont
  {Martin}}, \bibinfo {author} {\bibfnamefont {A.}~\bibnamefont {McClung}},
  \bibinfo {author} {\bibfnamefont {K.}~\bibnamefont {Choi}}, \bibinfo {author}
  {\bibfnamefont {D.~E.}\ \bibnamefont {Chang}}, \bibinfo {author}
  {\bibfnamefont {O.}~\bibnamefont {Painter}}, and\ \bibinfo {author}
  {\bibfnamefont {H.~J.}\ \bibnamefont {Kimble}},\ }\href {\doibase
  10.1038/ncomms4808} {\bibfield  {journal} {\bibinfo  {journal} {Nature
  Communications}\ }\textbf {\bibinfo {volume} {5}},\ \bibinfo
  {pages} {3808} (\bibinfo {year} {2014})}\BibitemShut {NoStop}%
\bibitem [{\citenamefont {Goban}\ \emph {et~al.}(2015)\citenamefont {Goban},
  \citenamefont {Hung}, \citenamefont {Hood}, \citenamefont {Yu}, \citenamefont
  {Muniz}, \citenamefont {Painter},and\ \citenamefont {Kimble}}]{GHH15}%
  \BibitemOpen
  \bibfield  {author} {\bibinfo {author} {\bibfnamefont {A.}~\bibnamefont
  {Goban}}, \bibinfo {author} {\bibfnamefont {C.-L.}\ \bibnamefont {Hung}},
  \bibinfo {author} {\bibfnamefont {J.~D.}\ \bibnamefont {Hood}}, \bibinfo
  {author} {\bibfnamefont {S.-P.}\ \bibnamefont {Yu}}, \bibinfo {author}
  {\bibfnamefont {J.~A.}\ \bibnamefont {Muniz}}, \bibinfo {author}
  {\bibfnamefont {O.}~\bibnamefont {Painter}}, and\ \bibinfo {author}
  {\bibfnamefont {H.~J.}\ \bibnamefont {Kimble}},\ }\href {\doibase
  10.1103/PhysRevLett.115.063601} {\bibfield  {journal} {\bibinfo  {journal}
  {Physical Review Letters}\ }\textbf {\bibinfo {volume} {115}},\ \bibinfo
  {pages} {063601} (\bibinfo {year} {2015})}\BibitemShut {NoStop}%
\bibitem [{\citenamefont {Vetsch}\ \emph {et~al.}(2010)\citenamefont {Vetsch},
  \citenamefont {Reitz}, \citenamefont {Sagu{\'{e}}}, \citenamefont {Schmidt},
  \citenamefont {Dawkins},and\ \citenamefont {Rauschenbeutel}}]{VRS10}%
  \BibitemOpen
  \bibfield  {author} {\bibinfo {author} {\bibfnamefont {E.}~\bibnamefont
  {Vetsch}}, \bibinfo {author} {\bibfnamefont {D.}~\bibnamefont {Reitz}},
  \bibinfo {author} {\bibfnamefont {G.}~\bibnamefont {Sagu{\'{e}}}}, \bibinfo
  {author} {\bibfnamefont {R.}~\bibnamefont {Schmidt}}, \bibinfo {author}
  {\bibfnamefont {S.~T.}\ \bibnamefont {Dawkins}}, and\ \bibinfo {author}
  {\bibfnamefont {A.}~\bibnamefont {Rauschenbeutel}},\ }\href {\doibase
  10.1103/PhysRevLett.104.203603} {\bibfield  {journal} {\bibinfo  {journal}
  {Physical Review Letters}\ }\textbf {\bibinfo {volume} {104}},\ \bibinfo
  {pages} {203603} (\bibinfo {year} {2010})}\BibitemShut {NoStop}%
\bibitem [{\citenamefont {Goban}\ \emph {et~al.}(2012)\citenamefont {Goban},
  \citenamefont {Choi}, \citenamefont {Alton}, \citenamefont {Ding},
  \citenamefont {Lacro{\^{u}}te}, \citenamefont {Pototschnig}, \citenamefont
  {Thiele}, \citenamefont {Stern},and\ \citenamefont {Kimble}}]{GCA12}%
  \BibitemOpen
  \bibfield  {author} {\bibinfo {author} {\bibfnamefont {A.}~\bibnamefont
  {Goban}}, \bibinfo {author} {\bibfnamefont {K.~S.}~\bibnamefont {Choi}},
  \bibinfo {author} {\bibfnamefont {D.~J.}~\bibnamefont {Alton}}, \bibinfo
  {author} {\bibfnamefont {D.}~\bibnamefont {Ding}}, \bibinfo {author}
  {\bibfnamefont {C.}~\bibnamefont {Lacro{\^{u}}te}}, \bibinfo {author}
  {\bibfnamefont {M.}~\bibnamefont {Pototschnig}}, \bibinfo {author}
  {\bibfnamefont {T.}~\bibnamefont {Thiele}}, \bibinfo {author} {\bibfnamefont
  {N.~P.}~\bibnamefont {Stern}}, and\ \bibinfo {author} {\bibfnamefont {H.~J.}\
  \bibnamefont {Kimble}},\ }\href {\doibase 10.1103/PhysRevLett.109.033603}
  {\bibfield  {journal} {\bibinfo  {journal} {Physical Review Letters}\
  }\textbf {\bibinfo {volume} {109}},\ \bibinfo {pages} {033603} (\bibinfo
  {year} {2012})}\BibitemShut {NoStop}%
\bibitem [{\citenamefont {Ghosh}\ \emph {et~al.}(2006)\citenamefont {Ghosh},
  \citenamefont {Bhagwat}, \citenamefont {Renshaw}, \citenamefont {Goh},
  \citenamefont {Gaeta},and\ \citenamefont {Kirby}}]{GBR06}%
  \BibitemOpen
  \bibfield  {author} {\bibinfo {author} {\bibfnamefont {S.}~\bibnamefont
  {Ghosh}}, \bibinfo {author} {\bibfnamefont {A.~R.}\ \bibnamefont {Bhagwat}},
  \bibinfo {author} {\bibfnamefont {C.~K.}\ \bibnamefont {Renshaw}}, \bibinfo
  {author} {\bibfnamefont {S.}~\bibnamefont {Goh}}, \bibinfo {author}
  {\bibfnamefont {A.~L.}\ \bibnamefont {Gaeta}}, and\ \bibinfo {author}
  {\bibfnamefont {B.~J.}\ \bibnamefont {Kirby}},\ }\href {\doibase
  10.1103/PhysRevLett.97.023603} {\bibfield  {journal} {\bibinfo  {journal}
  {Physical Review Letters}\ }\textbf {\bibinfo {volume} {97}},\ \bibinfo
  {pages} {023603} (\bibinfo {year} {2006})}\BibitemShut {NoStop}%
\bibitem [{\citenamefont {Christensen}\ \emph {et~al.}(2008)\citenamefont
  {Christensen}, \citenamefont {Will}, \citenamefont {Saba}, \citenamefont
  {Jo}, \citenamefont {Shin}, \citenamefont {Ketterle},and\ \citenamefont
  {Pritchard}}]{CWS08}%
  \BibitemOpen
  \bibfield  {author} {\bibinfo {author} {\bibfnamefont {C.~A.}~\bibnamefont
  {Christensen}}, \bibinfo {author} {\bibfnamefont {S.}~\bibnamefont {Will}},
  \bibinfo {author} {\bibfnamefont {M.}~\bibnamefont {Saba}}, \bibinfo {author}
  {\bibfnamefont {G.-B.}\ \bibnamefont {Jo}}, \bibinfo {author} {\bibfnamefont
  {Y.-I.}\ \bibnamefont {Shin}}, \bibinfo {author} {\bibfnamefont
  {W.}~\bibnamefont {Ketterle}}, and\ \bibinfo {author} {\bibfnamefont
  {D.}\ \bibnamefont {Pritchard}},\ }\href {\doibase
  10.1103/PhysRevA.78.033429} {\bibfield  {journal} {\bibinfo  {journal}
  {Physical Review A}\ }\textbf {\bibinfo {volume} {78}},\ \bibinfo {pages}
  {033429} (\bibinfo {year} {2008})}\BibitemShut {NoStop}%
\bibitem [{\citenamefont {Vorrath}\ \emph {et~al.}(2010)\citenamefont
  {Vorrath}, \citenamefont {M{\"{o}}ller}, \citenamefont {Windpassinger},
  \citenamefont {Bongs},and\ \citenamefont {Sengstock}}]{VMW10}%
  \BibitemOpen
  \bibfield  {author} {\bibinfo {author} {\bibfnamefont {S.}~\bibnamefont
  {Vorrath}}, \bibinfo {author} {\bibfnamefont {S.~A.}\ \bibnamefont
  {M{\"{o}}ller}}, \bibinfo {author} {\bibfnamefont {P.}~\bibnamefont
  {Windpassinger}}, \bibinfo {author} {\bibfnamefont {K.}~\bibnamefont
  {Bongs}}, and\ \bibinfo {author} {\bibfnamefont {K.}~\bibnamefont
  {Sengstock}},\ }\href {\doibase 10.1088/1367-2630/12/12/123015} {\bibfield
  {journal} {\bibinfo  {journal} {New Journal of Physics}\ }\textbf {\bibinfo
  {volume} {12}},\ \bibinfo {pages} {123015} (\bibinfo {year}
  {2010})}\BibitemShut {NoStop}%
\bibitem [{\citenamefont {Bajcsy}\ \emph {et~al.}(2011)\citenamefont {Bajcsy},
  \citenamefont {Hofferberth}, \citenamefont {Peyronel}, \citenamefont
  {Bali{\'{c}}}, \citenamefont {Liang}, \citenamefont {Zibrov}, \citenamefont
  {Vuleti{\'{c}}},and\ \citenamefont {Lukin}}]{BHP11}%
  \BibitemOpen
  \bibfield  {author} {\bibinfo {author} {\bibfnamefont {M.}~\bibnamefont
  {Bajcsy}}, \bibinfo {author} {\bibfnamefont {S.}~\bibnamefont {Hofferberth}},
  \bibinfo {author} {\bibfnamefont {T.}~\bibnamefont {Peyronel}}, \bibinfo
  {author} {\bibfnamefont {V.}~\bibnamefont {Bali{\'{c}}}}, \bibinfo {author}
  {\bibfnamefont {Q.}~\bibnamefont {Liang}}, \bibinfo {author} {\bibfnamefont
  {A.~S.}\ \bibnamefont {Zibrov}}, \bibinfo {author} {\bibfnamefont
  {V.}~\bibnamefont {Vuleti{\'{c}}}}, and\ \bibinfo {author} {\bibfnamefont
  {M.~D.}\ \bibnamefont {Lukin}},\ }\href {\doibase 10.1103/PhysRevA.83.063830}
  {\bibfield  {journal} {\bibinfo  {journal} {Physical Review A}\ }\textbf
  {\bibinfo {volume} {83}},\ \bibinfo {pages} {063830} (\bibinfo {year}
  {2011})}\BibitemShut {NoStop}%
\bibitem [{\citenamefont {Blatt}\ \emph {et~al.}(2014)\citenamefont {Blatt},
  \citenamefont {Halfmann},and\ \citenamefont {Peters}}]{BHP14}%
  \BibitemOpen
  \bibfield  {author} {\bibinfo {author} {\bibfnamefont {F.}~\bibnamefont
  {Blatt}}, \bibinfo {author} {\bibfnamefont {T.}~\bibnamefont {Halfmann}}, \
  and\ \bibinfo {author} {\bibfnamefont {T.}~\bibnamefont {Peters}},\ }\href
  {\doibase 10.1364/OL.39.000446} {\bibfield  {journal} {\bibinfo  {journal}
  {Optics Letters}\ }\textbf {\bibinfo {volume} {39}},\ \bibinfo {pages} {446}
  (\bibinfo {year} {2014})}\BibitemShut {NoStop}%
\bibitem [{\citenamefont {Kaczmarek}\ \emph {et~al.}(2015)\citenamefont
  {Kaczmarek}, \citenamefont {Saunders}, \citenamefont {Sprague}, \citenamefont
  {Kolthammer}, \citenamefont {Feizpour}, \citenamefont {Ledingham},
  \citenamefont {Brecht}, \citenamefont {Poem}, \citenamefont {Walmsley},and\
  \citenamefont {Nunn}}]{KSS15}%
  \BibitemOpen
  \bibfield  {author} {\bibinfo {author} {\bibfnamefont {K.~T.}\ \bibnamefont
  {Kaczmarek}}, \bibinfo {author} {\bibfnamefont {D.~J.}\ \bibnamefont
  {Saunders}}, \bibinfo {author} {\bibfnamefont {M.~R.}\ \bibnamefont
  {Sprague}}, \bibinfo {author} {\bibfnamefont {W.~S.}\ \bibnamefont
  {Kolthammer}}, \bibinfo {author} {\bibfnamefont {A.}~\bibnamefont
  {Feizpour}}, \bibinfo {author} {\bibfnamefont {P.~M.}\ \bibnamefont
  {Ledingham}}, \bibinfo {author} {\bibfnamefont {B.}~\bibnamefont {Brecht}},
  \bibinfo {author} {\bibfnamefont {E.}~\bibnamefont {Poem}}, \bibinfo {author}
  {\bibfnamefont {I.~A.}\ \bibnamefont {Walmsley}}, and\ \bibinfo {author}
  {\bibfnamefont {J.}~\bibnamefont {Nunn}},\ }\href {\doibase
  10.1364/OL.40.005582} {\bibfield  {journal} {\bibinfo  {journal} {Optics
  Letters}\ }\textbf {\bibinfo {volume} {40}},\ \bibinfo {pages} {5582}
  (\bibinfo {year} {2015})}\BibitemShut {NoStop}%
\bibitem [{\citenamefont {Schmidt} and\ \citenamefont
  {Imamoglu}(1996)}]{SI96}%
  \BibitemOpen
  \bibfield  {author} {\bibinfo {author} {\bibfnamefont {H.}~\bibnamefont
  {Schmidt}} and\ \bibinfo {author} {\bibfnamefont {A.}~\bibnamefont
  {Imamoglu}},\ }\href {\doibase 10.1364/OL.21.001936} {\bibfield  {journal}
  {\bibinfo  {journal} {Optics Letters}\ }\textbf {\bibinfo {volume} {21}},\
  \bibinfo {pages} {1936} (\bibinfo {year} {1996})}\BibitemShut {NoStop}%
\bibitem [{\citenamefont {Harris} and\ \citenamefont {Hau}(1999)}]{HH99}%
  \BibitemOpen
  \bibfield  {author} {\bibinfo {author} {\bibfnamefont {S.~E.}\ \bibnamefont
  {Harris}} and\ \bibinfo {author} {\bibfnamefont {L.~V.}\ \bibnamefont
  {Hau}},\ }\href {\doibase 10.1103/PhysRevLett.82.4611} {\bibfield  {journal}
  {\bibinfo  {journal} {Physical Review Letters}\ }\textbf {\bibinfo {volume}
  {82}},\ \bibinfo {pages} {4611} (\bibinfo {year} {1999})}\BibitemShut
  {NoStop}%
\bibitem [{\citenamefont {Fleischhauer} and\ \citenamefont
  {Lukin}(2000)}]{FL00}%
  \BibitemOpen
  \bibfield  {author} {\bibinfo {author} {\bibfnamefont {M.}~\bibnamefont
  {Fleischhauer}} and\ \bibinfo {author} {\bibfnamefont {M.~D.}\ \bibnamefont
  {Lukin}},\ }\href {\doibase 10.1103/PhysRevLett.84.5094} {\bibfield
  {journal} {\bibinfo  {journal} {Physical Review Letters}\ }\textbf {\bibinfo
  {volume} {84}},\ \bibinfo {pages} {5094} (\bibinfo {year}
  {2000})}\BibitemShut {NoStop}%
\bibitem [{\citenamefont {Fleischhauer} and\ \citenamefont
  {Lukin}(2002)}]{FL02}%
  \BibitemOpen
  \bibfield  {author} {\bibinfo {author} {\bibfnamefont {M.}~\bibnamefont
  {Fleischhauer}} and\ \bibinfo {author} {\bibfnamefont {M.~D.}\ \bibnamefont
  {Lukin}},\ }\href {\doibase 10.1103/PhysRevA.65.022314} {\bibfield  {journal}
  {\bibinfo  {journal} {Physical Review A}\ }\textbf {\bibinfo {volume} {65}},\
  \bibinfo {pages} {022314} (\bibinfo {year} {2002})}\BibitemShut {NoStop}%
\bibitem [{\citenamefont {Liu}\ \emph {et~al.}(2001)\citenamefont {Liu},
  \citenamefont {Dutton}, \citenamefont {Hau},and\ \citenamefont
  {Behroozi}}]{LDB01}%
  \BibitemOpen
  \bibfield  {author} {\bibinfo {author} {\bibfnamefont {C.}~\bibnamefont
  {Liu}}, \bibinfo {author} {\bibfnamefont {Z.}~\bibnamefont {Dutton}},
  \bibinfo {author} {\bibfnamefont {L.~V.}\ \bibnamefont {Hau}}, and\
  \bibinfo {author} {\bibfnamefont {C.~H.}\ \bibnamefont {Behroozi}},\ }\href
  {\doibase 10.1038/35054017} {\bibfield  {journal} {\bibinfo  {journal}
  {Nature}\ }\textbf {\bibinfo {volume} {409}},\ \bibinfo {pages} {490}
  (\bibinfo {year} {2001})}\BibitemShut {NoStop}%
\bibitem [{\citenamefont {Heinze}\ \emph {et~al.}(2013)\citenamefont {Heinze},
  \citenamefont {Hubrich},and\ \citenamefont {Halfmann}}]{HHH13}%
  \BibitemOpen
  \bibfield  {author} {\bibinfo {author} {\bibfnamefont {G.}~\bibnamefont
  {Heinze}}, \bibinfo {author} {\bibfnamefont {C.}~\bibnamefont {Hubrich}}, \
  and\ \bibinfo {author} {\bibfnamefont {T.}~\bibnamefont {Halfmann}},\ }\href
  {\doibase 10.1103/PhysRevLett.111.033601} {\bibfield  {journal} {\bibinfo
  {journal} {Physical Review Letters}\ }\textbf {\bibinfo {volume} {111}},\
  \bibinfo {pages} {033601} (\bibinfo {year} {2013})}\BibitemShut {NoStop}%
\bibitem [{\citenamefont {Lin}\ \emph {et~al.}(2009)\citenamefont {Lin},
  \citenamefont {Liao}, \citenamefont {Peters}, \citenamefont {Chou},
  \citenamefont {Wang}, \citenamefont {Cho}, \citenamefont {Kuan},and\
  \citenamefont {Yu}}]{LLP09}%
  \BibitemOpen
  \bibfield  {author} {\bibinfo {author} {\bibfnamefont {Y.-W.}\ \bibnamefont
  {Lin}}, \bibinfo {author} {\bibfnamefont {W.-T.}\ \bibnamefont {Liao}},
  \bibinfo {author} {\bibfnamefont {T.}~\bibnamefont {Peters}}, \bibinfo
  {author} {\bibfnamefont {H.-C.}\ \bibnamefont {Chou}}, \bibinfo {author}
  {\bibfnamefont {J.-S.}\ \bibnamefont {Wang}}, \bibinfo {author}
  {\bibfnamefont {H.-W.}\ \bibnamefont {Cho}}, \bibinfo {author} {\bibfnamefont
  {P.-C.}\ \bibnamefont {Kuan}}, and\ \bibinfo {author} {\bibfnamefont
  {I.~A.}\ \bibnamefont {Yu}},\ }\href {\doibase
  10.1103/PhysRevLett.102.213601} {\bibfield  {journal} {\bibinfo  {journal}
  {Physical Review Letters}\ }\textbf {\bibinfo {volume} {102}},\ \bibinfo
  {pages} {213601} (\bibinfo {year} {2009})}\BibitemShut {NoStop}%
\bibitem [{\citenamefont {Andr{\'{e}}}\ \emph {et~al.}(2005)\citenamefont
  {Andr{\'{e}}}, \citenamefont {Bajcsy}, \citenamefont {Lukin},and\
  \citenamefont {Zibrov}}]{ABZ05}%
  \BibitemOpen
  \bibfield  {author} {\bibinfo {author} {\bibfnamefont {A.}~\bibnamefont
  {Andr{\'{e}}}}, \bibinfo {author} {\bibfnamefont {M.}~\bibnamefont {Bajcsy}},
  \bibinfo {author} {\bibfnamefont {A.~S.}\ \bibnamefont {Zibrov}}, and\
  \bibinfo {author} {\bibfnamefont {M.~D.}\ \bibnamefont {Lukin}},\ }\href
  {\doibase 10.1103/PhysRevLett.94.063902} {\bibfield  {journal} {\bibinfo
  {journal} {Physical Review Letters}\ }\textbf {\bibinfo {volume} {94}},\
  \bibinfo {pages} {063902} (\bibinfo {year} {2005})}\BibitemShut {NoStop}%
\bibitem [{\citenamefont {Chen}\ \emph {et~al.}(2012)\citenamefont {Chen},
  \citenamefont {Lee}, \citenamefont {Hung}, \citenamefont {Chen},
  \citenamefont {Chen},and\ \citenamefont {Yu}}]{CLH12}%
  \BibitemOpen
  \bibfield  {author} {\bibinfo {author} {\bibfnamefont {Y.-H.}\ \bibnamefont
  {Chen}}, \bibinfo {author} {\bibfnamefont {M.-J.}\ \bibnamefont {Lee}},
  \bibinfo {author} {\bibfnamefont {W.}~\bibnamefont {Hung}}, \bibinfo {author}
  {\bibfnamefont {Y.-C.}\ \bibnamefont {Chen}}, \bibinfo {author}
  {\bibfnamefont {Y.-F.}\ \bibnamefont {Chen}}, and\ \bibinfo {author}
  {\bibfnamefont {I.~A.}\ \bibnamefont {Yu}},\ }\href {\doibase
  10.1103/PhysRevLett.108.173603} {\bibfield  {journal} {\bibinfo  {journal}
  {Physical Review Letters}\ }\textbf {\bibinfo {volume} {108}},\ \bibinfo
  {pages} {173603} (\bibinfo {year} {2012})}\BibitemShut {NoStop}%
\bibitem [{\citenamefont {Bajcsy}\ \emph {et~al.}(2009)\citenamefont {Bajcsy},
  \citenamefont {Hofferberth}, \citenamefont {Bali{\'{c}}}, \citenamefont
  {Peyronel}, \citenamefont {Hafezi}, \citenamefont {Zibrov}, \citenamefont
  {Vuleti{\'{c}}},and\ \citenamefont {Lukin}}]{BHB09}%
  \BibitemOpen
  \bibfield  {author} {\bibinfo {author} {\bibfnamefont {M.}~\bibnamefont
  {Bajcsy}}, \bibinfo {author} {\bibfnamefont {S.}~\bibnamefont {Hofferberth}},
  \bibinfo {author} {\bibfnamefont {V.}~\bibnamefont {Bali{\'{c}}}}, \bibinfo
  {author} {\bibfnamefont {T.}~\bibnamefont {Peyronel}}, \bibinfo {author}
  {\bibfnamefont {M.}~\bibnamefont {Hafezi}}, \bibinfo {author} {\bibfnamefont
  {A.~S.}\ \bibnamefont {Zibrov}}, \bibinfo {author} {\bibfnamefont
  {V.}~\bibnamefont {Vuleti{\'{c}}}}, and\ \bibinfo {author} {\bibfnamefont
  {M.~D.}\ \bibnamefont {Lukin}},\ }\href {\doibase
  10.1103/PhysRevLett.102.203902} {\bibfield  {journal} {\bibinfo  {journal}
  {Physical Review Letters}\ }\textbf {\bibinfo {volume} {102}},\ \bibinfo
  {pages} {203902} (\bibinfo {year} {2009})}\BibitemShut {NoStop}%
\bibitem [{\citenamefont {Grimm}\ \emph {et~al.}(2000)\citenamefont {Grimm},
  \citenamefont {Weidem{\"{u}}ller},and\ \citenamefont
  {Ovchinnikov}}]{GWO00}%
  \BibitemOpen
  \bibfield  {author} {\bibinfo {author} {\bibfnamefont {R.}~\bibnamefont
  {Grimm}}, \bibinfo {author} {\bibfnamefont {M.}~\bibnamefont
  {Weidem{\"{u}}ller}}, and\ \bibinfo {author} {\bibfnamefont {Y.~B.}\
  \bibnamefont {Ovchinnikov}},\ }\href {\doibase 10.1016/S1049-250X(08)60186-X}
  {\bibfield  {journal} {\bibinfo  {journal} {Advances In Atomic, Molecular,
  and Optical Physics}\ }\textbf {\bibinfo {volume} {42}},\ \bibinfo {pages}
  {95} (\bibinfo {year} {2000})}\BibitemShut {NoStop}%
\bibitem [{\citenamefont {Statkiewicz}\ \emph {et~al.}(2005)\citenamefont
  {Statkiewicz}, \citenamefont {Martynkien},and\ \citenamefont
  {Urbanczyk}}]{SMU05}%
  \BibitemOpen
  \bibfield  {author} {\bibinfo {author} {\bibfnamefont {G.}~\bibnamefont
  {Statkiewicz}}, \bibinfo {author} {\bibfnamefont {T.}~\bibnamefont
  {Martynkien}}, and\ \bibinfo {author} {\bibfnamefont {W.}~\bibnamefont
  {Urbanczyk}},\ }\href {\doibase 10.1016/j.optcom.2005.06.014} {\bibfield
  {journal} {\bibinfo  {journal} {Optics Communications}\ }\textbf {\bibinfo
  {volume} {255}},\ \bibinfo {pages} {175} (\bibinfo {year}
  {2005})}\BibitemShut {NoStop}%
\bibitem [{\citenamefont {Chen}\ \emph {et~al.}(2004)\citenamefont {Chen},
  \citenamefont {Li}, \citenamefont {Venkataraman}, \citenamefont {Gallagher},
  \citenamefont {Wood}, \citenamefont {Crowley}, \citenamefont {Carberry},
  \citenamefont {Zenteno},and\ \citenamefont {Koch}}]{CLV04}%
  \BibitemOpen
  \bibfield  {author} {\bibinfo {author} {\bibfnamefont {X.}~\bibnamefont
  {Chen}}, \bibinfo {author} {\bibfnamefont {M.-J.}\ \bibnamefont {Li}},
  \bibinfo {author} {\bibfnamefont {N.}~\bibnamefont {Venkataraman}}, \bibinfo
  {author} {\bibfnamefont {M.~T.}\ \bibnamefont {Gallagher}}, \bibinfo {author}
  {\bibfnamefont {W.~A.}\ \bibnamefont {Wood}}, \bibinfo {author}
  {\bibfnamefont {A.~M.}\ \bibnamefont {Crowley}}, \bibinfo {author}
  {\bibfnamefont {J.~P.}\ \bibnamefont {Carberry}}, \bibinfo {author}
  {\bibfnamefont {L.~A.}\ \bibnamefont {Zenteno}}, and\ \bibinfo {author}
  {\bibfnamefont {K.~W.}\ \bibnamefont {Koch}},\ }\href {\doibase
  10.1364/OPEX.12.003888} {\bibfield  {journal} {\bibinfo  {journal} {Optics
  Express}\ }\textbf {\bibinfo {volume} {12}},\ \bibinfo {pages} {3888}
  (\bibinfo {year} {2004})}\BibitemShut {NoStop}%
\bibitem [{\citenamefont {Bell}\ \emph {et~al.}(2007)\citenamefont {Bell},
  \citenamefont {Heywood}, \citenamefont {White}, \citenamefont {Close},and\
  \citenamefont {Scholten}}]{BHW07}%
  \BibitemOpen
  \bibfield  {author} {\bibinfo {author} {\bibfnamefont {S.~C.}\ \bibnamefont
  {Bell}}, \bibinfo {author} {\bibfnamefont {D.~M.}\ \bibnamefont {Heywood}},
  \bibinfo {author} {\bibfnamefont {J.~D.}\ \bibnamefont {White}}, \bibinfo
  {author} {\bibfnamefont {J.~D.}\ \bibnamefont {Close}}, and\ \bibinfo
  {author} {\bibfnamefont {R.~E.}\ \bibnamefont {Scholten}},\ }\href {\doibase
  10.1063/1.2734471} {\bibfield  {journal} {\bibinfo  {journal} {Applied
  Physics Letters}\ }\textbf {\bibinfo {volume} {90}},\ \bibinfo {pages}
  {171120} (\bibinfo {year} {2007})}\BibitemShut {NoStop}%
\bibitem [{\citenamefont {Palittapongarnpim}\ \emph {et~al.}(2012)\citenamefont
  {Palittapongarnpim}, \citenamefont {MacRae},and\ \citenamefont
  {Lvovsky}}]{PML12}%
  \BibitemOpen
  \bibfield  {author} {\bibinfo {author} {\bibfnamefont {P.}~\bibnamefont
  {Palittapongarnpim}}, \bibinfo {author} {\bibfnamefont {A.}~\bibnamefont
  {MacRae}}, and\ \bibinfo {author} {\bibfnamefont {A.~I.}\ \bibnamefont
  {Lvovsky}},\ }\href {\doibase 10.1063/1.4726458} {\bibfield  {journal}
  {\bibinfo  {journal} {Review of Scientific Instruments}\ }\textbf
  {\bibinfo {volume} {83}},\ \bibinfo {pages} {066101} (\bibinfo {year}
  {2012})}\BibitemShut {NoStop}%
\bibitem [{SM(2016)}]{SM}%
  \BibitemOpen
  {\emph {\bibinfo {title} {See Supplemental Material at
  http://journals.aps.org/prl/abstract/... for details on the experimental
  setup and the simulations.}}}\ (\bibinfo {year} {2016})\BibitemShut {NoStop}%
\bibitem [{\citenamefont {Steck}()}]{Steck2015}%
  \BibitemOpen
  \bibfield  {author} {\bibinfo {author} {\bibfnamefont {D.~A.}\ \bibnamefont
  {Steck}},\ }\href {http://steck.us/alkalidata} {\enquote {\bibinfo {title}
  {{Rubidium 87 D Line Data}},}\ \bibinfo {pages} 
	{available online at http://steck.us/alkalidata (revision 2.1.5, 13 January 2015)}}\BibitemShut {NoStop}%
\bibitem [{\citenamefont {Zimmer}\ \emph {et~al.}(2006)\citenamefont {Zimmer},
  \citenamefont {Andr{\'{e}}}, \citenamefont {Lukin},and\ \citenamefont
  {Fleischhauer}}]{ZAL06}%
  \BibitemOpen
  \bibfield  {author} {\bibinfo {author} {\bibfnamefont {F.~E.}\ \bibnamefont
  {Zimmer}}, \bibinfo {author} {\bibfnamefont {A.}~\bibnamefont {Andr{\'{e}}}},
  \bibinfo {author} {\bibfnamefont {M.~D.}\ \bibnamefont {Lukin}}, and\
  \bibinfo {author} {\bibfnamefont {M.}~\bibnamefont {Fleischhauer}},\ }\href
  {\doibase 10.1016/j.optcom.2006.03.075} {\bibfield  {journal} {\bibinfo
  {journal} {Optics Communications}\ }\textbf {\bibinfo {volume} {264}},\
  \bibinfo {pages} {441} (\bibinfo {year} {2006})}\BibitemShut {NoStop}%
\bibitem [{\citenamefont {Wu}\ \emph {et~al.}(2010)\citenamefont {Wu},
  \citenamefont {Artoni},and\ \citenamefont {{La Rocca}}}]{WAL10b}%
  \BibitemOpen
  \bibfield  {author} {\bibinfo {author} {\bibfnamefont {J.-H.}\ \bibnamefont
  {Wu}}, \bibinfo {author} {\bibfnamefont {M.}~\bibnamefont {Artoni}}, and\
  \bibinfo {author} {\bibfnamefont {G.~C.}\ \bibnamefont {{La Rocca}}},\ }\href
  {\doibase 10.1103/PhysRevA.82.013807} {\bibfield  {journal} {\bibinfo
  {journal} {Physical Review A}\ }\textbf {\bibinfo {volume} {82}},\ \bibinfo
  {pages} {013807} (\bibinfo {year} {2010})}\BibitemShut {NoStop}%
\bibitem [{\citenamefont {Peters}\ \emph
  {et~al.}(2012{\natexlab{a}})\citenamefont {Peters}, \citenamefont {Su},
  \citenamefont {Chen}, \citenamefont {Wang}, \citenamefont {Gou},and\
  \citenamefont {Yu}}]{PSC12}%
  \BibitemOpen
  \bibfield  {author} {\bibinfo {author} {\bibfnamefont {T.}~\bibnamefont
  {Peters}}, \bibinfo {author} {\bibfnamefont {S.-W.}\ \bibnamefont {Su}},
  \bibinfo {author} {\bibfnamefont {Y.-H.}\ \bibnamefont {Chen}}, \bibinfo
  {author} {\bibfnamefont {J.-S.}\ \bibnamefont {Wang}}, \bibinfo {author}
  {\bibfnamefont {S.-C.}\ \bibnamefont {Gou}}, and\ \bibinfo {author}
  {\bibfnamefont {I.~A.}\ \bibnamefont {Yu}},\ }\href {\doibase
  10.1103/PhysRevA.85.023838} {\bibfield  {journal} {\bibinfo  {journal}
  {Physical Review A}\ }\textbf {\bibinfo {volume} {85}},\ \bibinfo {pages}
  {023838} (\bibinfo {year} {2012}{\natexlab{a}})}\BibitemShut {NoStop}%
\bibitem [{\citenamefont {Peters}\ \emph
  {et~al.}(2012{\natexlab{b}})\citenamefont {Peters}, \citenamefont {Wittrock},
  \citenamefont {Blatt}, \citenamefont {Halfmann},and\ \citenamefont
  {Yatsenko}}]{PWB12}%
  \BibitemOpen
  \bibfield  {author} {\bibinfo {author} {\bibfnamefont {T.}~\bibnamefont
  {Peters}}, \bibinfo {author} {\bibfnamefont {B.}~\bibnamefont {Wittrock}},
  \bibinfo {author} {\bibfnamefont {F.}~\bibnamefont {Blatt}}, \bibinfo
  {author} {\bibfnamefont {T.}~\bibnamefont {Halfmann}}, and\ \bibinfo
  {author} {\bibfnamefont {L.~P.}~\bibnamefont {Yatsenko}},\ }\href {\doibase
  10.1103/PhysRevA.85.063416} {\bibfield  {journal} {\bibinfo  {journal}
  {Physical Review A}\ }\textbf {\bibinfo {volume} {85}},\ \bibinfo {pages}
  {063416} (\bibinfo {year} {2012}{\natexlab{b}})}\BibitemShut {NoStop}%
\bibitem [{\citenamefont {Sagle}\ \emph {et~al.}(1996)\citenamefont {Sagle},
  \citenamefont {Namiotka},and\ \citenamefont {Huennekens}}]{SNH96}%
  \BibitemOpen
  \bibfield  {author} {\bibinfo {author} {\bibfnamefont {J.}~\bibnamefont
  {Sagle}}, \bibinfo {author} {\bibfnamefont {R.~K.}\ \bibnamefont {Namiotka}},
  and\ \bibinfo {author} {\bibfnamefont {J.}~\bibnamefont {Huennekens}},\
  }\href {\doibase 10.1088/0953-4075/29/12/023} {\bibfield  {journal} {\bibinfo
   {journal} {Journal of Physics B: Atomic, Molecular and Optical Physics}\
  }\textbf {\bibinfo {volume} {29}},\ \bibinfo {pages} {2629} (\bibinfo {year}
  {1996})}\BibitemShut {NoStop}%
\bibitem [{\citenamefont {Peters}\ \emph {et~al.}(2009)\citenamefont {Peters},
  \citenamefont {Chen}, \citenamefont {Wang}, \citenamefont {Lin},and\
  \citenamefont {Yu}}]{PCW09}%
  \BibitemOpen
  \bibfield  {author} {\bibinfo {author} {\bibfnamefont {T.}~\bibnamefont
  {Peters}}, \bibinfo {author} {\bibfnamefont {Y.-H.}\ \bibnamefont {Chen}},
  \bibinfo {author} {\bibfnamefont {J.-S.}\ \bibnamefont {Wang}}, \bibinfo
  {author} {\bibfnamefont {Y.-W.}\ \bibnamefont {Lin}}, and\ \bibinfo
  {author} {\bibfnamefont {I.~A.}\ \bibnamefont {Yu}},\ }\href {\doibase
  10.1364/OE.17.006665} {\bibfield  {journal} {\bibinfo  {journal} {Optics
  Express}\ }\textbf {\bibinfo {volume} {17}},\ \bibinfo {pages} {6665}
  (\bibinfo {year} {2009})}\BibitemShut {NoStop}%
\bibitem [{\citenamefont {Gouraud}\ \emph {et~al.}(2015)\citenamefont
  {Gouraud}, \citenamefont {Maxein}, \citenamefont {Nicolas}, \citenamefont
  {Morin},and\ \citenamefont {Laurat}}]{GMD15}%
  \BibitemOpen
  \bibfield  {author} {\bibinfo {author} {\bibfnamefont {B.}~\bibnamefont
  {Gouraud}}, \bibinfo {author} {\bibfnamefont {D.}~\bibnamefont {Maxein}},
  \bibinfo {author} {\bibfnamefont {A.}~\bibnamefont {Nicolas}}, \bibinfo
  {author} {\bibfnamefont {O.}~\bibnamefont {Morin}}, and\ \bibinfo {author}
  {\bibfnamefont {J.}~\bibnamefont {Laurat}},\ }\href {\doibase
  10.1103/PhysRevLett.114.180503} {\bibfield  {journal} {\bibinfo  {journal}
  {Physical Review Letters}\ }\textbf {\bibinfo {volume} {114}},\ \bibinfo
  {pages} {180503} (\bibinfo {year} {2015})}\BibitemShut {NoStop}%
\bibitem [{\citenamefont {Sayrin}\ \emph {et~al.}(2015)\citenamefont {Sayrin},
  \citenamefont {Clausen}, \citenamefont {Albrecht}, \citenamefont
  {Schneeweiss},and\ \citenamefont {Rauschenbeutel}}]{SCA15}%
  \BibitemOpen
  \bibfield  {author} {\bibinfo {author} {\bibfnamefont {C.}~\bibnamefont
  {Sayrin}}, \bibinfo {author} {\bibfnamefont {C.}~\bibnamefont {Clausen}},
  \bibinfo {author} {\bibfnamefont {B.}~\bibnamefont {Albrecht}}, \bibinfo
  {author} {\bibfnamefont {P.}~\bibnamefont {Schneeweiss}}, and\ \bibinfo
  {author} {\bibfnamefont {A.}~\bibnamefont {Rauschenbeutel}},\ }\href
  {\doibase 10.1364/OPTICA.2.000353} {\bibfield  {journal} {\bibinfo  {journal}
  {Optica}\ }\textbf {\bibinfo {volume} {2}},\ \bibinfo {pages} {353} (\bibinfo
  {year} {2015})}\BibitemShut {NoStop}%
\bibitem [{\citenamefont {Winoto}\ \emph {et~al.}(1999)\citenamefont {Winoto},
  \citenamefont {DePue}, \citenamefont {Bramall},and\ \citenamefont
  {Weiss}}]{WDB99}%
  \BibitemOpen
  \bibfield  {author} {\bibinfo {author} {\bibfnamefont {S.~L.}~\bibnamefont
  {Winoto}}, \bibinfo {author} {\bibfnamefont {M.~T.}~\bibnamefont {DePue}},
  \bibinfo {author} {\bibfnamefont {N.~E.}~\bibnamefont {Bramall}}, and\
  \bibinfo {author} {\bibfnamefont {D.~S.}\ \bibnamefont {Weiss}},\ }\href
  {\doibase 10.1103/PhysRevA.59.R19} {\bibfield  {journal} {\bibinfo  {journal}
  {Physical Review A}\ }\textbf {\bibinfo {volume} {59}},\ \bibinfo {pages}
  {R19} (\bibinfo {year} {1999})}\BibitemShut {NoStop}%
\bibitem [{\citenamefont {Dudin}\ \emph {et~al.}(2013)\citenamefont {Dudin},
  \citenamefont {Li},and\ \citenamefont {Kuzmich}}]{DLK13}%
  \BibitemOpen
  \bibfield  {author} {\bibinfo {author} {\bibfnamefont {Y.~O.}\ \bibnamefont
  {Dudin}}, \bibinfo {author} {\bibfnamefont {L.}~\bibnamefont {Li}}, and\
  \bibinfo {author} {\bibfnamefont {A.}~\bibnamefont {Kuzmich}},\ }\href
  {\doibase 10.1103/PhysRevA.87.031801} {\bibfield  {journal} {\bibinfo
  {journal} {Physical Review A}\ }\textbf {\bibinfo {volume} {87}},\ \bibinfo
  {pages} {031801} (\bibinfo {year} {2013})}\BibitemShut {NoStop}%
\bibitem [{\citenamefont {Peyronel}\ \emph
  {et~al.}(2012{\natexlab{b}})\citenamefont {Peyronel}, \citenamefont {Bajcsy},
  \citenamefont {Hofferberth}, \citenamefont {Balic}, \citenamefont {Hafezi},
  \citenamefont {{Qiyu Liang}}, \citenamefont {Zibrov}, \citenamefont
  {Vuleti{\'{c}}},and\ \citenamefont {Lukin}}]{PBH12}%
  \BibitemOpen
  \bibfield  {author} {\bibinfo {author} {\bibfnamefont {T.}~\bibnamefont
  {Peyronel}}, \bibinfo {author} {\bibfnamefont {M.}~\bibnamefont {Bajcsy}},
  \bibinfo {author} {\bibfnamefont {S.}~\bibnamefont {Hofferberth}}, \bibinfo
  {author} {\bibfnamefont {V.}~\bibnamefont {Balic}}, \bibinfo {author}
  {\bibfnamefont {M.}~\bibnamefont {Hafezi}}, \bibinfo {author} {\bibnamefont
  {{Qiyu Liang}}}, \bibinfo {author} {\bibfnamefont {A.}~\bibnamefont
  {Zibrov}}, \bibinfo {author} {\bibfnamefont {V.}~\bibnamefont
  {Vuleti{\'{c}}}}, and\ \bibinfo {author} {\bibfnamefont {M.~D.}\
  \bibnamefont {Lukin}},\ }\href {\doibase 10.1109/JSTQE.2012.2196414}
  {\bibfield  {journal} {\bibinfo  {journal} {IEEE Journal of Selected Topics
  in Quantum Electronics}\ }\textbf {\bibinfo {volume} {18}},\ \bibinfo {pages}
  {1747} (\bibinfo {year} {2012}{\natexlab{b}})}\BibitemShut {NoStop}%
\end{thebibliography}

\begin{thebibliography}{16}%
\makeatletter
\providecommand \@ifxundefined [1]{%
 \@ifx{#1\undefined}
}%
\providecommand \@ifnum [1]{%
 \ifnum #1\expandafter \@firstoftwo
 \else \expandafter \@secondoftwo
 \fi
}%
\providecommand \@ifx [1]{%
 \ifx #1\expandafter \@firstoftwo
 \else \expandafter \@secondoftwo
 \fi
}%
\providecommand \natexlab [1]{#1}%
\providecommand \enquote  [1]{``#1''}%
\providecommand \bibnamefont  [1]{#1}%
\providecommand \bibfnamefont [1]{#1}%
\providecommand \citenamefont [1]{#1}%
\providecommand \href@noop [0]{\@secondoftwo}%
\providecommand \href [0]{\begingroup \@sanitize@url \@href}%
\providecommand \@href[1]{\@@startlink{#1}\@@href}%
\providecommand \@@href[1]{\endgroup#1\@@endlink}%
\providecommand \@sanitize@url [0]{\catcode `\\12\catcode `\$12\catcode
  `\&12\catcode `\#12\catcode `\^12\catcode `\_12\catcode `\%12\relax}%
\providecommand \@@startlink[1]{}%
\providecommand \@@endlink[0]{}%
\providecommand \url  [0]{\begingroup\@sanitize@url \@url }%
\providecommand \@url [1]{\endgroup\@href {#1}{\urlprefix }}%
\providecommand \urlprefix  [0]{URL }%
\providecommand \Eprint [0]{\href }%
\providecommand \doibase [0]{http://dx.doi.org/}%
\providecommand \selectlanguage [0]{\@gobble}%
\providecommand \bibinfo  [0]{\@secondoftwo}%
\providecommand \bibfield  [0]{\@secondoftwo}%
\providecommand \translation [1]{[#1]}%
\providecommand \BibitemOpen [0]{}%
\providecommand \bibitemStop [0]{}%
\providecommand \bibitemNoStop [0]{.\EOS\space}%
\providecommand \EOS [0]{\spacefactor3000\relax}%
\providecommand \BibitemShut  [1]{\csname bibitem#1\endcsname}%
\let\auto@bib@innerbib\@empty
\bibitem [{\citenamefont {Lin}\ \emph {et~al.}(2009)\citenamefont {Lin},
  \citenamefont {Liao}, \citenamefont {Peters}, \citenamefont {Chou},
  \citenamefont {Wang}, \citenamefont {Cho}, \citenamefont {Kuan},and\
  \citenamefont {Yu}}]{LLP09SM}%
  \BibitemOpen
  \bibfield  {author} {\bibinfo {author} {\bibfnamefont {Y.-W.}\ \bibnamefont
  {Lin}}, \bibinfo {author} {\bibfnamefont {W.-T.}\ \bibnamefont {Liao}},
  \bibinfo {author} {\bibfnamefont {T.}~\bibnamefont {Peters}}, \bibinfo
  {author} {\bibfnamefont {H.-C.}\ \bibnamefont {Chou}}, \bibinfo {author}
  {\bibfnamefont {J.-S.}\ \bibnamefont {Wang}}, \bibinfo {author}
  {\bibfnamefont {H.-W.}\ \bibnamefont {Cho}}, \bibinfo {author} {\bibfnamefont
  {P.-C.}\ \bibnamefont {Kuan}}, and\ \bibinfo {author} {\bibfnamefont
  {I.~A.}\ \bibnamefont {Yu}},\ }\href {\doibase
  10.1103/PhysRevLett.102.213601} {\bibfield  {journal} {\bibinfo  {journal}
  {Physical Review Letters}\ }\textbf {\bibinfo {volume} {102}},\ \bibinfo
  {pages} {213601} (\bibinfo {year} {2009})}\BibitemShut {NoStop}%
\bibitem [{\citenamefont {Wu}\ \emph {et~al.}(2010)\citenamefont {Wu},
  \citenamefont {Artoni},and\ \citenamefont {{La Rocca}}}]{WAL10bSM}%
  \BibitemOpen
  \bibfield  {author} {\bibinfo {author} {\bibfnamefont {J.-H.}\ \bibnamefont
  {Wu}}, \bibinfo {author} {\bibfnamefont {M.}~\bibnamefont {Artoni}}, and\
  \bibinfo {author} {\bibfnamefont {G.~C.}\ \bibnamefont {{La Rocca}}},\ }\href
  {\doibase 10.1103/PhysRevA.82.013807} {\bibfield  {journal} {\bibinfo
  {journal} {Physical Review A}\ }\textbf {\bibinfo {volume} {82}},\ \bibinfo
  {pages} {013807} (\bibinfo {year} {2010})}\BibitemShut {NoStop}%
\bibitem [{\citenamefont {Peters}\ \emph
  {et~al.}(2012{\natexlab{a}})\citenamefont {Peters}, \citenamefont {Su},
  \citenamefont {Chen}, \citenamefont {Wang}, \citenamefont {Gou},and\
  \citenamefont {Yu}}]{PSC12SM}%
  \BibitemOpen
  \bibfield  {author} {\bibinfo {author} {\bibfnamefont {T.}~\bibnamefont
  {Peters}}, \bibinfo {author} {\bibfnamefont {S.-W.}\ \bibnamefont {Su}},
  \bibinfo {author} {\bibfnamefont {Y.-H.}\ \bibnamefont {Chen}}, \bibinfo
  {author} {\bibfnamefont {J.-S.}\ \bibnamefont {Wang}}, \bibinfo {author}
  {\bibfnamefont {S.-C.}\ \bibnamefont {Gou}}, and\ \bibinfo {author}
  {\bibfnamefont {I.~A.}\ \bibnamefont {Yu}},\ }\href {\doibase
  10.1103/PhysRevA.85.023838} {\bibfield  {journal} {\bibinfo  {journal}
  {Physical Review A}\ }\textbf {\bibinfo {volume} {85}},\ \bibinfo {pages}
  {023838} (\bibinfo {year} {2012}{\natexlab{a}})}\BibitemShut {NoStop}%
\bibitem [{\citenamefont {Steck}()}]{Steck2015SM}%
  \BibitemOpen
  \bibfield  {author} {\bibinfo {author} {\bibfnamefont {D.~A.}\ \bibnamefont
  {Steck}},\ }\href {http://steck.us/alkalidata} {\enquote {\bibinfo {title}
  {{Rubidium 87 D Line Data}},}\ \bibinfo {pages} 
	{available online at http://steck.us/alkalidata (revision 2.1.5, 13 January 2015)}}\BibitemShut {NoStop}%
\bibitem [{\citenamefont {Zimmer}\ \emph {et~al.}(2006)\citenamefont {Zimmer},
  \citenamefont {Andr{\'{e}}}, \citenamefont {Lukin},and\ \citenamefont
  {Fleischhauer}}]{ZAL06SM}%
  \BibitemOpen
  \bibfield  {author} {\bibinfo {author} {\bibfnamefont {F.~E.}\ \bibnamefont
  {Zimmer}}, \bibinfo {author} {\bibfnamefont {A.}~\bibnamefont {Andr{\'{e}}}},
  \bibinfo {author} {\bibfnamefont {M.~D.}\ \bibnamefont {Lukin}}, and\
  \bibinfo {author} {\bibfnamefont {M.}~\bibnamefont {Fleischhauer}},\ }\href
  {\doibase 10.1016/j.optcom.2006.03.075} {\bibfield  {journal} {\bibinfo
  {journal} {Optics Communications}\ }\textbf {\bibinfo {volume} {264}},\
  \bibinfo {pages} {441} (\bibinfo {year} {2006})}\BibitemShut {NoStop}%
\bibitem [{\citenamefont {Sagle}\ \emph {et~al.}(1996)\citenamefont {Sagle},
  \citenamefont {Namiotka},and\ \citenamefont {Huennekens}}]{SNH96SM}%
  \BibitemOpen
  \bibfield  {author} {\bibinfo {author} {\bibfnamefont {J.}~\bibnamefont
  {Sagle}}, \bibinfo {author} {\bibfnamefont {R.~K.}\ \bibnamefont {Namiotka}},
  and\ \bibinfo {author} {\bibfnamefont {J.}~\bibnamefont {Huennekens}},\
  }\href {\doibase 10.1088/0953-4075/29/12/023} {\bibfield  {journal} {\bibinfo
   {journal} {Journal of Physics B: Atomic, Molecular and Optical Physics}\
  }\textbf {\bibinfo {volume} {29}},\ \bibinfo {pages} {2629} (\bibinfo {year}
  {1996})}\BibitemShut {NoStop}%
\bibitem [{\citenamefont {Grimm}\ \emph {et~al.}(2000)\citenamefont {Grimm},
  \citenamefont {Weidem{\"{u}}ller},and\ \citenamefont
  {Ovchinnikov}}]{GWO00SM}%
  \BibitemOpen
  \bibfield  {author} {\bibinfo {author} {\bibfnamefont {R.}~\bibnamefont
  {Grimm}}, \bibinfo {author} {\bibfnamefont {M.}~\bibnamefont
  {Weidem{\"{u}}ller}}, and\ \bibinfo {author} {\bibfnamefont {Y.~B.}\
  \bibnamefont {Ovchinnikov}},\ }\href {\doibase 10.1016/S1049-250X(08)60186-X}
  {\bibfield  {journal} {\bibinfo  {journal} {Advances In Atomic, Molecular,
  and Optical Physics}\ }\textbf {\bibinfo {volume} {42}},\ \bibinfo {pages}
  {95} (\bibinfo {year} {2000})}\BibitemShut {NoStop}%
\bibitem [{\citenamefont {Peters}\ \emph
  {et~al.}(2012{\natexlab{b}})\citenamefont {Peters}, \citenamefont {Wittrock},
  \citenamefont {Blatt}, \citenamefont {Halfmann},and\ \citenamefont
  {Yatsenko}}]{PWB12SM}%
  \BibitemOpen
  \bibfield  {author} {\bibinfo {author} {\bibfnamefont {T.}~\bibnamefont
  {Peters}}, \bibinfo {author} {\bibfnamefont {B.}~\bibnamefont {Wittrock}},
  \bibinfo {author} {\bibfnamefont {F.}~\bibnamefont {Blatt}}, \bibinfo
  {author} {\bibfnamefont {T.}~\bibnamefont {Halfmann}}, and\ \bibinfo
  {author} {\bibfnamefont {L.~P.}~\bibnamefont {Yatsenko}},\ }\href {\doibase
  10.1103/PhysRevA.85.063416} {\bibfield  {journal} {\bibinfo  {journal}
  {Physical Review A}\ }\textbf {\bibinfo {volume} {85}},\ \bibinfo {pages}
  {063416} (\bibinfo {year} {2012}{\natexlab{b}})}\BibitemShut {NoStop}%
\bibitem [{\citenamefont {Ketterle}\ \emph {et~al.}(1993)\citenamefont
  {Ketterle}, \citenamefont {Davis}, \citenamefont {Joffe}, \citenamefont
  {Martin},and\ \citenamefont {Pritchard}}]{KDJ93SM}%
  \BibitemOpen
  \bibfield  {author} {\bibinfo {author} {\bibfnamefont {W.}~\bibnamefont
  {Ketterle}}, \bibinfo {author} {\bibfnamefont {K.}~\bibnamefont {Davis}},
  \bibinfo {author} {\bibfnamefont {M.}~\bibnamefont {Joffe}}, \bibinfo
  {author} {\bibfnamefont {A.}~\bibnamefont {Martin}}, and\ \bibinfo {author}
  {\bibfnamefont {D.~E.}\ \bibnamefont {Pritchard}},\ }\href {\doibase
  10.1103/PhysRevLett.70.2253} {\bibfield  {journal} {\bibinfo  {journal}
  {Physical Review Letters}\ }\textbf {\bibinfo {volume} {70}},\ \bibinfo
  {pages} {2253} (\bibinfo {year} {1993})}\BibitemShut {NoStop}%
\bibitem [{\citenamefont {Blatt}\ \emph {et~al.}(2014)\citenamefont {Blatt},
  \citenamefont {Halfmann},and\ \citenamefont {Peters}}]{BHP14SM}%
  \BibitemOpen
  \bibfield  {author} {\bibinfo {author} {\bibfnamefont {F.}~\bibnamefont
  {Blatt}}, \bibinfo {author} {\bibfnamefont {T.}~\bibnamefont {Halfmann}}, \
  and\ \bibinfo {author} {\bibfnamefont {T.}~\bibnamefont {Peters}},\ }\href
  {\doibase 10.1364/OL.39.000446} {\bibfield  {journal} {\bibinfo  {journal}
  {Optics Letters}\ }\textbf {\bibinfo {volume} {39}},\ \bibinfo {pages} {446}
  (\bibinfo {year} {2014})}\BibitemShut {NoStop}%
\bibitem [{\citenamefont {Peters}\ \emph {et~al.}(2009)\citenamefont {Peters},
  \citenamefont {Chen}, \citenamefont {Wang}, \citenamefont {Lin},and\
  \citenamefont {Yu}}]{PCW09SM}%
  \BibitemOpen
  \bibfield  {author} {\bibinfo {author} {\bibfnamefont {T.}~\bibnamefont
  {Peters}}, \bibinfo {author} {\bibfnamefont {Y.-H.}\ \bibnamefont {Chen}},
  \bibinfo {author} {\bibfnamefont {J.-S.}\ \bibnamefont {Wang}}, \bibinfo
  {author} {\bibfnamefont {Y.-W.}\ \bibnamefont {Lin}}, and\ \bibinfo
  {author} {\bibfnamefont {I.~A.}\ \bibnamefont {Yu}},\ }\href {\doibase
  10.1364/OE.17.006665} {\bibfield  {journal} {\bibinfo  {journal} {Optics
  Express}\ }\textbf {\bibinfo {volume} {17}},\ \bibinfo {pages} {6665}
  (\bibinfo {year} {2009})}\BibitemShut {NoStop}%
\bibitem [{\citenamefont {Gouraud}\ \emph {et~al.}(2015)\citenamefont
  {Gouraud}, \citenamefont {Maxein}, \citenamefont {Nicolas}, \citenamefont
  {Morin},and\ \citenamefont {Laurat}}]{GMD15SM}%
  \BibitemOpen
  \bibfield  {author} {\bibinfo {author} {\bibfnamefont {B.}~\bibnamefont
  {Gouraud}}, \bibinfo {author} {\bibfnamefont {D.}~\bibnamefont {Maxein}},
  \bibinfo {author} {\bibfnamefont {A.}~\bibnamefont {Nicolas}}, \bibinfo
  {author} {\bibfnamefont {O.}~\bibnamefont {Morin}}, and\ \bibinfo {author}
  {\bibfnamefont {J.}~\bibnamefont {Laurat}},\ }\href {\doibase
  10.1103/PhysRevLett.114.180503} {\bibfield  {journal} {\bibinfo  {journal}
  {Physical Review Letters}\ }\textbf {\bibinfo {volume} {114}},\ \bibinfo
  {pages} {180503} (\bibinfo {year} {2015})}\BibitemShut {NoStop}%
\bibitem [{\citenamefont {Statkiewicz}\ \emph {et~al.}(2005)\citenamefont
  {Statkiewicz}, \citenamefont {Martynkien},and\ \citenamefont
  {Urbanczyk}}]{SMU05SM}%
  \BibitemOpen
  \bibfield  {author} {\bibinfo {author} {\bibfnamefont {G.}~\bibnamefont
  {Statkiewicz}}, \bibinfo {author} {\bibfnamefont {T.}~\bibnamefont
  {Martynkien}}, and\ \bibinfo {author} {\bibfnamefont {W.}~\bibnamefont
  {Urbanczyk}},\ }\href {\doibase 10.1016/j.optcom.2005.06.014} {\bibfield
  {journal} {\bibinfo  {journal} {Optics Communications}\ }\textbf {\bibinfo
  {volume} {255}},\ \bibinfo {pages} {175} (\bibinfo {year}
  {2005})}\BibitemShut {NoStop}%
\bibitem [{\citenamefont {Chen}\ \emph {et~al.}(2004)\citenamefont {Chen},
  \citenamefont {Li}, \citenamefont {Venkataraman}, \citenamefont {Gallagher},
  \citenamefont {Wood}, \citenamefont {Crowley}, \citenamefont {Carberry},
  \citenamefont {Zenteno},and\ \citenamefont {Koch}}]{CLV04SM}%
  \BibitemOpen
  \bibfield  {author} {\bibinfo {author} {\bibfnamefont {X.}~\bibnamefont
  {Chen}}, \bibinfo {author} {\bibfnamefont {M.-J.}\ \bibnamefont {Li}},
  \bibinfo {author} {\bibfnamefont {N.}~\bibnamefont {Venkataraman}}, \bibinfo
  {author} {\bibfnamefont {M.~T.}\ \bibnamefont {Gallagher}}, \bibinfo {author}
  {\bibfnamefont {W.~A.}\ \bibnamefont {Wood}}, \bibinfo {author}
  {\bibfnamefont {A.~M.}\ \bibnamefont {Crowley}}, \bibinfo {author}
  {\bibfnamefont {J.~P.}\ \bibnamefont {Carberry}}, \bibinfo {author}
  {\bibfnamefont {L.~A.}\ \bibnamefont {Zenteno}}, and\ \bibinfo {author}
  {\bibfnamefont {K.~W.}\ \bibnamefont {Koch}},\ }\href {\doibase
  10.1364/OPEX.12.003888} {\bibfield  {journal} {\bibinfo  {journal} {Optics
  Express}\ }\textbf {\bibinfo {volume} {12}},\ \bibinfo {pages} {3888}
  (\bibinfo {year} {2004})}\BibitemShut {NoStop}%
\bibitem [{\citenamefont {Bajcsy}\ \emph {et~al.}(2011)\citenamefont {Bajcsy},
  \citenamefont {Hofferberth}, \citenamefont {Peyronel}, \citenamefont
  {Bali{\'{c}}}, \citenamefont {Liang}, \citenamefont {Zibrov}, \citenamefont
  {Vuleti{\'{c}}},and\ \citenamefont {Lukin}}]{BHP11SM}%
  \BibitemOpen
  \bibfield  {author} {\bibinfo {author} {\bibfnamefont {M.}~\bibnamefont
  {Bajcsy}}, \bibinfo {author} {\bibfnamefont {S.}~\bibnamefont {Hofferberth}},
  \bibinfo {author} {\bibfnamefont {T.}~\bibnamefont {Peyronel}}, \bibinfo
  {author} {\bibfnamefont {V.}~\bibnamefont {Bali{\'{c}}}}, \bibinfo {author}
  {\bibfnamefont {Q.}~\bibnamefont {Liang}}, \bibinfo {author} {\bibfnamefont
  {A.~S.}\ \bibnamefont {Zibrov}}, \bibinfo {author} {\bibfnamefont
  {V.}~\bibnamefont {Vuleti{\'{c}}}}, and\ \bibinfo {author} {\bibfnamefont
  {M.~D.}\ \bibnamefont {Lukin}},\ }\href {\doibase 10.1103/PhysRevA.83.063830}
  {\bibfield  {journal} {\bibinfo  {journal} {Physical Review A}\ }\textbf
  {\bibinfo {volume} {83}},\ \bibinfo {pages} {063830} (\bibinfo {year}
  {2011})}\BibitemShut {NoStop}%
\bibitem [{\citenamefont {Palittapongarnpim}\ \emph {et~al.}(2012)\citenamefont
  {Palittapongarnpim}, \citenamefont {MacRae},and\ \citenamefont
  {Lvovsky}}]{PML12SM}%
  \BibitemOpen
  \bibfield  {author} {\bibinfo {author} {\bibfnamefont {P.}~\bibnamefont
  {Palittapongarnpim}}, \bibinfo {author} {\bibfnamefont {A.}~\bibnamefont
  {MacRae}}, and\ \bibinfo {author} {\bibfnamefont {A.~I.}\ \bibnamefont
  {Lvovsky}},\ }\href {\doibase 10.1063/1.4726458} {\bibfield  {journal}
  {\bibinfo  {journal} {Review of Scientific Instruments}\ }\textbf
  {\bibinfo {volume} {83}},\ \bibinfo {pages} {066101} (\bibinfo {year}
  {2012})}\BibitemShut {NoStop}%
\end{thebibliography}
\end{document}